\documentclass[aps,prd,superscriptaddress,showkeys,showpacs,twocolumn,a4paper]{revtex4-1}
\usepackage{ulem,amssymb}
\usepackage{cancel}
\usepackage{color}
\usepackage{graphicx}
\usepackage{epsfig}
\usepackage{mathrsfs}
\usepackage{amsmath,amsthm}
\usepackage[dvipsnames]{xcolor}
\usepackage{float}
\usepackage[bottom]{footmisc}
\definecolor{darkblue}{RGB}{0,0,196}
\definecolor{darkgreen}{RGB}{0,120,0}
\usepackage[colorlinks=true,linktocpage=true,linkcolor=darkblue,citecolor=red,urlcolor=darkblue]{hyperref}

\newcommand{\be}{\begin{equation}}
\newcommand{\ee}{\end{equation}}
\newcommand{\ba}{\begin{eqnarray}}
\newcommand{\ea}{\end{eqnarray}}

\begin{document}
\title{Collective excitations of a hot QCD medium in  a time dependent background magnetic field}
\author{Gowthama K K}
\email{k$_$gowthama@iitgn.ac.in}
\affiliation{Indian Institute of Technology Gandhinagar, Gandhinagar-382355, Gujarat, India}

\author{Vinod Chandra}
\email{vchandra@iitgn.ac.in}
\affiliation{Indian Institute of Technology Gandhinagar, Gandhinagar-382355, Gujarat, India}

\begin{abstract}
{Collective modes within a hot Quantum Chromodynamics (QCD) medium are obtained from the polarization tensor, considering both constant and time-varying electromagnetic fields. In both scenarios, five complex modes emerge, reliant on the wave vector ($k$), with electrical conductivity exerting significant influence. 
The impact of the modes on the energy loss of heavy quarks in the hot QCD medium with a background electromagnetic field has been studied by obtaining the induced electric field in terms of the polarization tensor while invoking Wong's equations. The findings are seen to be consistent with analogous approaches, reinforcing the significance of the results.
}
\end{abstract}

\maketitle

 \section{Introduction}
The Relativistic Heavy Ion Collider (RHIC) and the Large Hadron Collider (LHC) have conducted Heavy-Ion experiments, providing evidence of the existence of an intriguing form of matter known as Quark-Gluon Plasma (QGP)~\cite{Adams:2005dq}. This unique state of matter arises from the intense interactions of quarks and gluons as dictated by the underlying theory of strong interaction force, {\it viz.}, Quantum Chromodynamics (QCD)  exists for very short times (a few fm/c) at exceedingly high temperatures. Given the complex nature of the strong interaction and, hence, the QGP, its direct investigation proves challenging, necessitating indirect methods to unravel its properties. Among these, two prominent signatures are jet quenching and collective flow, which have been observed at the  RHIC and the LHC. Through these experiments, compelling evidence has emerged substantiating the presence of the QGP and its strong coupling nature that is exemplified by the observation of an extremely low shear viscosity to entropy density ratio, further supporting the notion of the QGP as a near-ideal fluid~\cite{Ryu:2015vwa,Denicol:2015nhu}.

Experiments at the RHIC and the LHC observed enhanced directed flow of $D/D^{0}$, indicating the presence of a  strong initial magnetic field in heavy-ion collisions~\cite{Acharya:2019ijj,Adam:2019wnk}. The produced magnetic field decays slowly in the medium and may persist for a longer time due to the back reaction from the medium. The behaviour of the magnetic field in the medium depends on the conductivity of the medium~\cite{Tuchin:2013apa,McLerran:2013hla, Stewart:2021mjz,Tuchin:2019gkg,Yan:2021zjc}. The behaviour of the QGP medium may depend on the persistent field and hence is vital to study the response of the medium to such a field. Numerous studies delve into this, encompassing electrical conductivity in weak and time dependent electromagnetic fields~\cite{Feng:2017tsh,Thakur:2019bnf,Dey:2019axu,Hattori:2016lqx,Fukushima:2017lvb, Dash:2020vxk,Kurian:2017yxj,Kurian:2019fty,Ghosh:2019ubc, Gowthama:2020ghl,K:2021sct}, as well as thermal and thermoelectric response of the medium~\cite{Denicol:2012vq,Kapusta:2012zb,Kurian:2021zyb,K:2022pzc}.

The investigation of the QGP necessitates an in-depth exploration of its collective excitations, which carry crucial information into both equilibrated and evolving non(near)-equilibrated QGP. In this context, it is imperative to examine both quark and gluonic collective modes within the QGP medium. The collective modes of the QGP medium have been studied through the linear response formalism~\cite{Carrington:2003je,Schenke:2006xu}, with momentum anisotropy\cite{Carrington:2014bla,Romatschke:2003ms,Romatschke:2004jh}, with equation of state effects~\cite{Kumar:2017bja,Jamal:2017dqs} and with an external electromagnetic field~\cite{Formanek:2021blc}. 

In this study, we focus on comprehending the collective excitations of the QGP medium in the presence of a weak, time-varying magnetic field through the polarization tensor. This marks a novel attempt to understand the properties of the hot QCD medium under the influence of time-dependent magnetic fields, with the inclusion of electric ($\sigma_e$) and Hall ($\sigma_H$) conductivities. Our analysis unfolds the presence of two complex modes and three purely imaginary modes, two of which display positive growth, signifying potential instabilities within the medium.

Heavy quarks/anti-quarks, produced in the early stages of collisions, offer valuable insights into the entire space-time evolution of the QGP medium, making them excellent probes of the medium. The energy loss of heavy quarks has been studied through the transport theory approach~\cite{Elias:2014hua,Han:2017nfz,YousufJamal:2019pen,Ghosh:2023ghi,Jamal:2023ncn} and within the finite temperature field theory approach~\cite{PhysRevD.44.R2625,Mrowczynski:1991da,Thoma:1991jum}. Our focus here is on studying how bottom and charm quarks lose energy while traversing an isotropic, collisional QGP. We model the collisions employing the RTA collisional kernel within the Boltzmann transport equation. When a charged quark moves through the hot QCD medium, it induces an electric field, which, in turn, generates a Lorentz force back on the quark, causing it to lose energy. The induced electric field is obtained by solving Maxwell's equation with the polarization tensor obtained through the transport theory. The induced field is employed in Wong's equation to study the energy loss of the heavy quarks.

The paper is organized as follows. Section II deals with the basic formalism of polarization tensor with consideration of different cases of the magnetic field. In Section III, the brief mathematical formalism for the polarization tensor and the different collective modes are presented in different sub-sections. The formalism for energy loss of heavy quarks has been presented in Section IV. Section V contains the results and discussion, and Section VI offers a summary and conclusions of the present work.

{\it Notations and conventions:} The subscript $k$ denotes the particle species. The quantity $q_{k}$ is the electric charge of the $k$th species. The particle velocity is defined as ${\bf v}=\frac{{\bf p}}{\epsilon}$, where ${\bf p}$ is the momentum and $\epsilon=\sqrt{p^2+m_f^2}$ is the energy (with $m_f$ as the mass of quark with flavor $f$) of the particle. The component of a three vector ${\bf A}$ is denoted with the Latin indices $A^i$. The quantities  $E=|{\bf E}|$ and $B=|{\bf B}|$ denote the magnitude of the electric and magnetic fields.
\section{Polarization Tensor in hot QCD Medium in Electro-magnetic fields }\label{II}
The propagator of the hot QCD medium can be obtained from the polarization tensor, invoking Maxwell's equation. The polarization tensor is related to the induced current in the medium due to the external electromagnetic field. In this work, we adopt the transport theory approach while employing the  Relaxation-Time Approximation (RTA) for the collisional kernel.

We consider the transport equation with $F^{\mu \nu}$, the external electromagnetic field tensor, 
\begin{align}\label{1.1}
&{v}^{\mu}\partial_{\mu} \delta f_k+q_k v_{\mu}F^{\mu \nu}\partial^{(p)}_{\nu} f_k=-\frac{\delta f_k}{\tau_R},
\end{align}
with $q_k$ being the charge of the particle and $f_k$ being the quarks/antiquarks momentum distribution function.
\begin{align}\label{1.2}
&f_k = f^{0}_k +\delta f_k, &&f^0_k=\frac{1}{1+\exp{\big(\beta( \epsilon_k \mp \mu)\big)}},
\end{align}
with $f^{0}_k$ being the near equilibrium distribution function. The collision term has been chosen to be the RTA kernel, with $\tau_R$ being the relaxation time~\cite{Hosoya:1983xm}.

The general form of the induced vector current in the QCD medium with a non-vanishing quark chemical potential $\mu$ in terms of quark and antiquark momentum distribution function $f_k=f^0_k+\delta f_k$ is as follows,
\begin{align}\label{1.3}
{\bf j}&=2N_c\sum_f \int d P\,{\bf v}\,\Big(q_q f_q-q_{\bar{q}}f_{\bar{q}}\Big),  
 \end{align}
where $v_i$ is the component of velocity and $d P = \frac{d^3 \textbf{p}}{(2\pi)^3}$.
The flavour summation (over the up, down, and strange quarks) arises from the degeneracy factor $2N_c\sum_f$ of the quarks/antiquarks with $N_c$ number of colours.\\
The electric current can be written as, 
\begin{equation}
   j^{i} = \sigma_e \delta^{ij}E_{j} +\sigma_{H} \epsilon^{ij}E_{j} 
\end{equation}

where $\sigma_e$ and $\sigma_H$ are the Ohmic and Hall conductivities, respectively. The conductivities can be obtained by solving the transport equation with the non-equilibrium part of the distribution function $\delta f_k=({\bf{p}}.{\bf \Xi} ) \frac{\partial f^0_k}{\partial \epsilon_k},$ containing the electric and magnetic fields. The vector ${\bf{\Xi}}$ is related to the strength of the electromagnetic field and its first-order (leading order) spacetime derivatives, with the following form,
\begin{align}\label{1.4}
\mathbf{\Xi} =& \alpha_1\textbf{E}+ \alpha_2\dot{\textbf{E}}+ \alpha_3(\textbf{E}\times \textbf{B})+ \alpha_4(\dot{\textbf{E}}\times \textbf{B})+ \alpha_5(\textbf{E}\times \dot{\textbf{B}})\nonumber\\&+\alpha_6 ({\pmb \nabla} \times \textbf{E}) +\alpha_7 \textbf{B}+\alpha_8 \dot{\textbf{B}}+\alpha_9 ({\pmb \nabla} \times \textbf{B}).
\end{align} 
Here, $\alpha_{i}$ ($i=(1, 2,.., 9)$) are the unknown functions that relate to the respective electric charge transport coefficients and can be obtained by the microscopic description of the QCD medium. The coefficients are found, as derived in Ref.~\cite{K:2021sct}, to be,
\begin{align}\label{20}
 &\alpha_1 =-\frac{\Tilde{\Omega}_k}{2}(I_1 e^{\eta_1} +I_2 e^{\eta_2} ),\\
 &\alpha_3 =\frac{q_{fk}i}{2\epsilon}(I_1 e^{\eta_1} -I_2 e^{\eta_2} ),\label{20.2}\\
 &\alpha_5 =-\frac{\tau_R q_{fk}i}{2\epsilon}(I_1 e^{\eta_1} -I_2 e^{\eta_2}  ),\label{20.4}\\
 &\alpha_2 =\frac{(\frac{\Tilde{\Omega}_k\tau_R}{2} +\frac{i\Omega^2_k\tau^2_R}{2})I_1 e^{\eta_1}+(\frac{\Tilde{\Omega}_k\tau_R}{2}-\frac{i\Omega^2_k\tau^2_R}{2}) I_2 e^{\eta_2}}{1+\Omega^2_k\tau^2_R}\label{20.1}\\
 &\alpha_4 =\frac{-(\frac{\Tilde{\Omega}^2_k\tau^2_R}{2F} +\frac{iq_{fk}\tau_R}{2\epsilon})I_1 e^{\eta_1}-(\frac{\Tilde{\Omega}^2_k\tau^2_R}{2F}-\frac{iq_{fk}\tau_R}{2\epsilon}) I_2 e^{\eta_2}}{1+\Omega^2_k\tau^2_R},\label{20.3}
\end{align}
with $\Omega_k=\frac{q_{f\, k}B}{\epsilon}$ represents the cyclotron frequency at finite $B$ and $\Tilde{\Omega}_k=\frac{q_{f\,k}F}{\epsilon}$, where $F = \sqrt{B(B-\tau_R \dot{B})}$ for the time-varying magnetic field.
Here the terms $\eta_j$ and $I_j$ are defined as, 
\begin{align}\label{21}
    &\eta_j = -\frac{t}{\tau_R} +a_j\frac{q_{f_k} i}{\epsilon}\int F dt,
    &&I_j = \int \frac{ e^{-\eta_j}}{F},
\end{align}
with $a_1=1$ and $a_2=-1$. Now, we solve the master equation of $\alpha_i$ for various choices of the electromagnetic fields \textbf{E} and \textbf{B}.

\subsubsection*{Case I: Constant electric and magnetic fields}
For the case of a constant electric and magnetic field, the terms $\alpha_2,\alpha_4,\alpha_5$ disappear, and the coefficients $\alpha_1$ and $\alpha_3$ correspond to the Ohmic and Hall conductivity respectfully as described in Ref.~\cite{Gowthama:2020ghl}. The expression for $\alpha_1$ and $\alpha_3$ are given by, 
\begin{align}\label{1.5}
 &\alpha_1 = \frac{-\epsilon q_{f_k}}{\tau_R [(\frac{\epsilon}{\tau_R})^2 +(q_{f_k} B)^2]},
 &&\alpha_3 =\frac{-q_{f_k}^2 }{[(\frac{\epsilon}{\tau_R})^2 +(q_{f_k} B)^2]}.
\end{align}
Employing Eq.~(\ref{1.5}) in Eq.~(\ref{1.3}), we obtain  $j^i=\sigma_{e} \delta^{ij} E_j +\sigma_H \epsilon^{ij}E_j$ with $\sigma_{e}$ and $\sigma_H$ denote the electrical and Hall conductivites, respectively. The results obtained are in agreement with the observations of Ref.~\cite{Feng:2017tsh}. The chemical potential plays an important role with the Hall conductivity vanishing at $\mu =0$, zero chemical potential. The electric current is written as,
\begin{align}\label{18}
j_{e}=&\frac{E(t)}{3} 2N_c\sum_k \sum_f (q_{f_k})^2 \int \frac{d^3 \textbf{p}}{(2\pi)^3} p^2(-\frac{\partial f^0_k}{\partial \epsilon})\nonumber\\&\times  \frac{1}{\tau_R \big[(\frac{\epsilon}{\tau_R})^2 +(q_{f_k} B)^2\big]},
\end{align}
\begin{align}\label{18.3}
j_{H}=&\frac{E(t)}{3}2N_c \sum_k \sum_f (q_{f_k})^2 \int \frac{d^3 \textbf{p}}{(2\pi)^3} \frac{p^2}{\epsilon} (-\frac{\partial f^0_k}{\partial \epsilon}) \nonumber\\&\times \frac{q_{f_k} B}{\big[(\frac{\epsilon}{\tau_R})^2 +(q_{f_k} B)^2\big]},
\end{align}
which agrees with the results of~\cite{Puglisi:2014sha}. The transverse component, $\sigma_H$, vanishes in the presence of a strong magnetic field due to the $1+1-$D Landau dynamic, with the longitudinal electrical conductivity being the dominant electrical charge transport.  
\subsubsection*{Case II: Response to time-varying electromagnetic field}

\begin{figure*}
    \vspace{-3cm}
    \centering
    \hspace{-2.5cm}
    \includegraphics[width=0.571\textwidth]{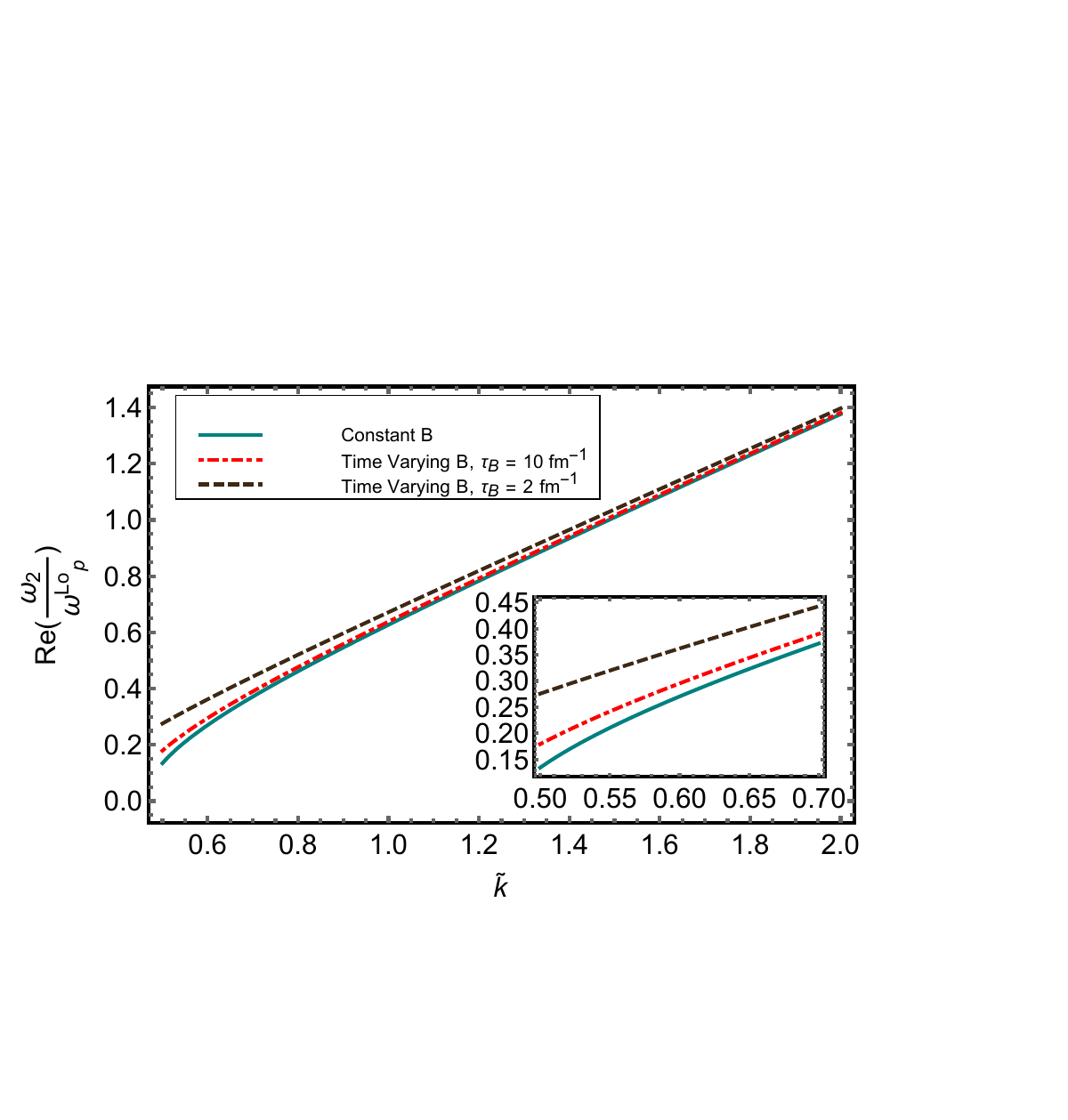}
    \hspace{-0.5cm}
    \includegraphics[width=0.571\textwidth]{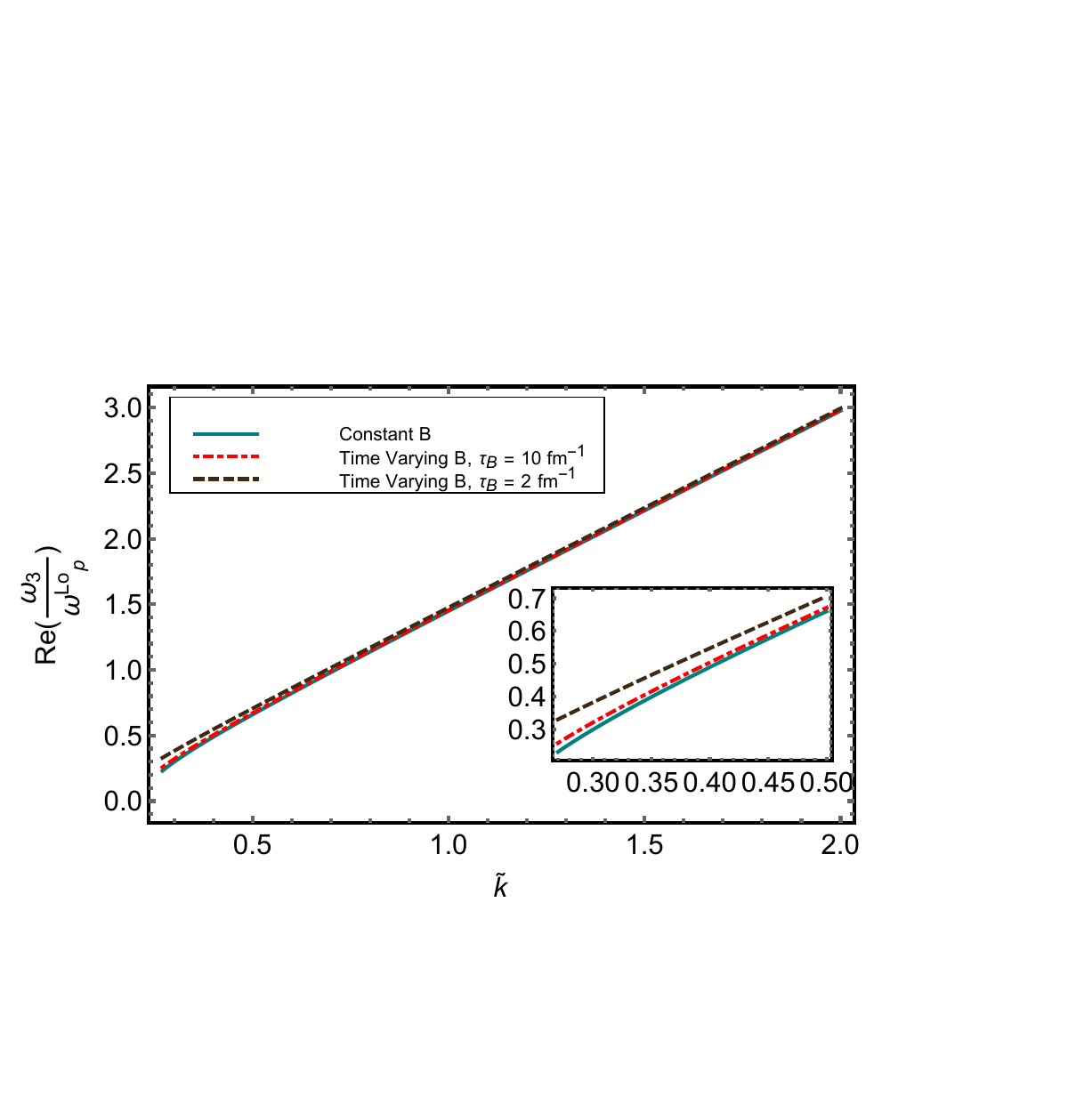}
    \hspace{-2.5cm}
    \vspace{-2.5cm}
    \caption{\small Dispersion relation curve of $\omega_2$ (Left Panel) and $\omega_3$ (Right Panel) modes with different conditions of the magnetic field, constant magnetic field and, time varying magnetic field with $\tau_b = 10,2$ $fm^{-1}$. The graph is plotted at $T = 2T_C$ with $T_C = 0.17$ GeV.}
\label{f1}
\end{figure*}

In the case of time dependent electric and magnetic fields, we choose a particular magnetic field of the form $\textbf{B} = B_0 e^{-\frac{t}{\tau_B}}\hat{\bf z}$, where $B_0$ is its amplitude and $\tau_B$ is the decay time parameter~\cite{Hongo:2013cqa, Satow:2014lia} such that $F = B\sqrt{1+\frac{\tau_R}{\tau_B}}$. With the assumption that the cyclotron frequency $\Omega_k$ is approximately equal to the decay frequency ($\tau_B^{-1}$) of the magnetic field, Eq.~(\ref{21}) reduces to the following form,
\begin{align}\label{30}
    &\eta_j = -\frac{t}{\tau_R} +a_j\Bigg(i\frac{\sqrt{1+\frac{\tau_R}{\tau_B}}}{\tau_B}t\Bigg),
\end{align}
\begin{align}
    &I_j = \frac{1}{\sqrt{1+\frac{\tau_R}{\tau_B}} B_0} \frac{e^{\Big(\frac{1}{\tau_R} +\frac{1}{\tau_B}-a_ji\frac{\sqrt{1+\frac{\tau_R}{\tau_B}}}{\tau_B} \Big)t}}{\Big( \frac{1}{\tau_R} +\frac{1}{\tau_B}-a_ji\frac{\sqrt{1+\frac{\tau_R}{\tau_B}}}{\tau_B} \Big)}.\label{30.2}
\end{align}
we proceed with the estimation of all $\alpha_i$ coefficients by substituting Eq.~(\ref{30}) and Eq.~(\ref{30.2}) in Eq.~(\ref{20})-(\ref{20.3}). Incorporating the non-zero contributions associated with $\alpha_i, (i=1,2,..5)$ in Eq.~(\ref{1.4}), we obtain five components of the induced current ${\bf j}=j_e{\bf\hat{e}}+j_H({\bf\hat{e}}\times{\bf\hat{b}})$ as follows,
\begin{align}\label{1.8}
    &j_e = j_e^{(0)}+j_e^{(1)}, &&j_H = j_H^{(0)}+j_H^{(1)}+j_H^{(2)}, 
\end{align}
where $j_e$ corresponds to the electric current in the direction of the electric field ${\bf\hat{e}}$ and $j_H$ is the electrical current in the direction perpendicular to both electric and magnetic fields $({\bf\hat{e}}\times{\bf\hat{b}})$ with,
\begin{align}\label{1.9}
    &j_e^{(0)} =  \frac{2E}{3} N_c \sum_k \sum_f (q_{fk})^2 \int dP\frac{ p^2}{\epsilon^2}(-\frac{\partial f_k^0}{\partial \epsilon}) {M}_1,\\ 
    &j_e^{(1)} =\frac{2\dot{E}}{3} N_c \sum_k \sum_f (q_{fk})^2\int dP \frac{ p^2}{\epsilon^2}\frac{\partial f_k^0}{\partial \epsilon}M_2,\label{31.1} \\
     &j_H^{(0)} = \frac{2E}{3} N_c \sum_k \sum_f (q_{fk})^3\int dP \frac{ p^2}{\epsilon^3}(-\frac{\partial f_k^0}{\partial \epsilon}) M,\label{31.2}\\ 
     &j_H^{(1)} =\frac{2\dot{E}}{3} N_c \sum_k \sum_f (q_{fk})^3\int dP \frac{ p^2}{\epsilon^3}\frac{\partial f_k^0}{\partial \epsilon}M_3,\label{31.3}\\ 
     &j_H^{(2)} = \frac{2E}{3\tau_B} N_c \sum_k \sum_f (q_{fk})^3\int dP \frac{ p^2}{\epsilon^3}(-\frac{\partial f_k^0}{\partial \epsilon}) \tau_R M,\label{31.4}
\end{align}
where $\dot{E}=|{\bf \dot{E}}|$ and $M_j (j=1, 2, 3)$ functions can be defined as $M_1 =\big(\frac{1}{\tau_R} +\frac{1}{\tau_B}\big) M$, $M_2=-\big(\tau_R M_1-\frac{\tau_R^2}{\tau_B^2}M\big)/\big(1+(\frac{\tau_R}{\tau_B})^2\big)$ and $M_3=\big(\tau_R M + \tau_R^2M_1 \big)/\big(1+(\frac{\tau_R}{\tau_B})^2\big)$ with, 
\begin{align}\label{1.10}
 M = \Bigg[{ \frac{1}{ \tau_R} +\frac{1}{\tau_B}+\frac{\sqrt{1+\frac{\tau_R}{\tau_B}}}{\tau_B} }\Bigg]^{-1}.
\end{align}
In  Eqs.~(\ref{1.9}) to (\ref{31.4}), the leading-order Ohmic current is represented by $j_e^{(0)}$, while $j_e^{(1)}$ signifies the correction due to the Ohmic current resulting from the time-varying nature of the fields. The Hall current in the medium, produced by perpendicular electric and magnetic fields, is denoted by $j_H^{(0)}$, while its corrections from ($\dot{\textbf{E}} \times \textbf{B}$) and ($\textbf{E} \times \dot{\textbf{B}}$) are represented by $j_H^{(1)}$ and $j_H^{(2)}$ respectfully.

From these expressions for the induced current, the polarization tensor can be obtained using $\Pi^{\mu \nu} = \frac{\delta j^{\mu}_{ind}}{\delta A_{\nu}}$ in the temporal gauge with $A_0 =0$ and $A^i = \frac{E^i}{i\omega}$,
\begin{equation}\label{1.11}
    \Pi^{i j} = i\omega \sigma_e \delta^{ij} +i \omega \sigma_H \epsilon^{ij}.
\end{equation}
The Fourier transformed Maxwell's equation is, 
\begin{equation}\label{1.12}
    -ik_{\nu} F^{\mu \nu} (K) = J^{\mu}_{ind}(K)+J^{\mu}_{ext}(K),
\end{equation}
where $J^{\mu}_{ext}(K)$ is the external current. The induced current can be expressed in terms of the polarization tensor, written as, 
\begin{equation}\label{1.13}
   J^{\mu}_{ind}(K) = \Pi^{\mu \nu} (K) A_{\nu}.
\end{equation}
The Maxwell equation can be written as,
\begin{equation}\label{1.14}
    [K^2 g^{\mu \nu} -k^{\mu}k^{\nu} +\Pi^{\mu \nu}]A_{\nu}(K) =J^{\mu}_{ext}(K).
\end{equation}
Considering the temporal gauge $A_0 =0$, writing the equation in terms of the electric field,
\begin{equation}\label{1.15}
    [\Delta^{ij}(k)]^{-1} E_j = i\omega J^i_{ext},
\end{equation}
where, 
\begin{equation}\label{1.16}
   [\Delta^{ij}(k)]^{-1} = [(k^2-\omega^2)\delta^{ij} -k^ik^j +\Pi^{ij}(k)],
\end{equation}
is the inverse of the propagator. The poles of the propagator, $[\Delta^{ij}(k)]$, give us the dispersion relation of the collective modes. 
\section{Collective modes: Finding the Poles of the Propagator}\label{III}

\subsection{Decomposition of Polarization Tensor}
The polarization tensor encodes the interaction of the medium. Hence, the properties of the QGP medium can be analyzed by studying the structure of the polarization tensor. We have obtained the form of the polarization tensor through the semi-classical transport theory approach in Eq.~(\ref{1.11}). The polarization tensor in the isotropic medium can be expanded in terms of two components. The longitudinal component, $P^{ij}_L = k^i k^j/k^2$ and the transverse component $P^{ij}_T = \delta^{ij} - k^i k^j/k^2$. The background magnetic field induces anisotropy in the medium, and hence an additional two components are required to describe the polarization tensor in the presence of a magnetic field, $P^{ij}_b = b^i b^j$ and $P^{ij}_{bk} = \epsilon^{ij\eta}b_{\eta}$, where $b^i = B^i/|B|$ is the unit vector along the magnetic field and is taken to be such that, $b^ik_i = 0$. With this, the polarization tensor can be expanded as, 
\begin{equation}\label{1.17}
    \Pi^{ij} = \alpha_0 P^{ij}_L +\beta_0 P^{ij}_T +\gamma_0 P^{ij}_b +\delta_0 P^{ij}_{bk},
\end{equation}
where $\alpha, \beta, \gamma, \delta$ are the scalar structure constants. These can be obtained by taking the appropriate projections of the polarization tensor with the following projections,
\begin{align}\label{1.18}
    &P_L .P_L = (k^i k^j/k^2)(k_i k_j/k^2) =1,\\ 
    & P_T.P_T = (\delta^{ij} - k^i k^j/k^2)(\delta_{ij} - k_i k_j/k^2) = 2,\\
    & P_L. P_T = (k^i k^j/k^2).(\delta_{ij} - k_i k_j/k^2)=0,\\ 
    &P_L.P_b = (k^i k^j/k^2)(b_i b_j)=0,\\ 
    &P_T.P_b = (\delta^{ij} - k^i k^j/k^2)(b_i b_j) = 1,\\
    &P_L.P_{bk} = (k^i k^j/k^2)(\epsilon_{ij\eta} b^{\eta}) =0,\\ 
    &P_T.P_{bk} = (\delta^{ij} - k^i k^j/k^2)(\epsilon_{ij\eta} b^{\eta}) = 0,\\ 
    &P_b. P_{bk} = (b^i b^j)(\epsilon_{ij\eta} b^{\eta}) = 0,\\ 
    &P_{bk} .P_{bk} = (\epsilon^{ij\eta} b_{\eta})(\epsilon_{ij\eta} b^{\eta}) = 1,\label{1.81}
\end{align}
where, $P_{a}.P_{c}=(P_a)^{ij} (P_{c})_{ij}$, the dot product refers to the contraction of both indices. The polarization tensor in Eq.~(\ref{1.11}) can be written in terms of the projection tensors as, 
\begin{equation}
    \Pi^{ij} = i\omega \sigma_e P_{L}+ i\omega \sigma_e P_{T}+i\omega \sigma_H P_{bk}.
\end{equation}
Using the contractions, Eqs.~(\ref{1.18})-(\ref{1.81}), to find the coefficients of Eq.~(\ref{1.17}) in terms of Eq.~(\ref{1.11}),
\begin{align}\label{1.19}
    &\Pi.P_L = \alpha_0 = i\omega \sigma_e, &&\Pi.P_T = 2\beta_0 +\gamma_0 =2i\omega \sigma_e,\\
    &\Pi.P_b = \gamma_0 + \beta_0 = i\omega \sigma_e, && \Pi.P_{bk} = \delta_0 = i\omega \sigma_H.
\end{align}
With these, the coefficients are found to be, 
\begin{align}\label{1.20}
    &\alpha_0 = i\omega \sigma_e, &&\beta_0 = i\omega \sigma_e,\\ \nonumber
    &\gamma_0 = 0, &&\delta_0 = i\omega \sigma_H.
\end{align}

\begin{figure*}
    \vspace{-1cm}
    \centering
    \hspace{-2.5cm}
    \includegraphics[width=0.571\textwidth]{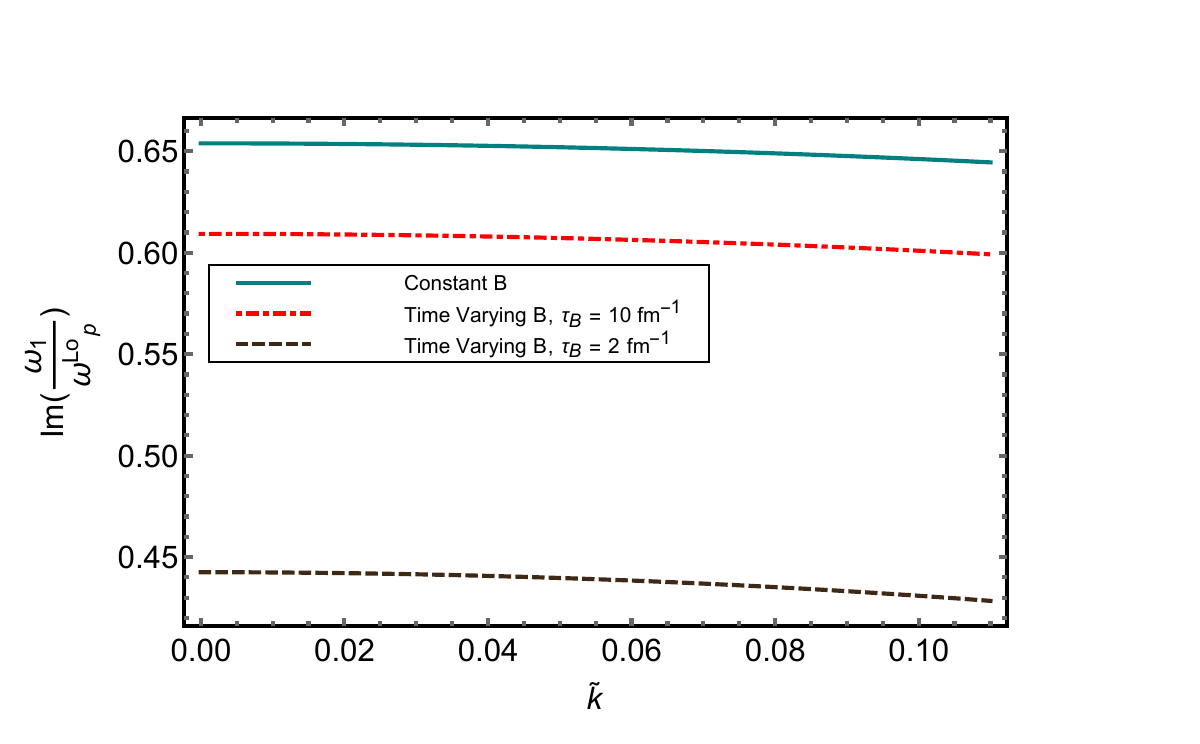}
    \hspace{-.5cm}
    \includegraphics[width=0.58\textwidth]{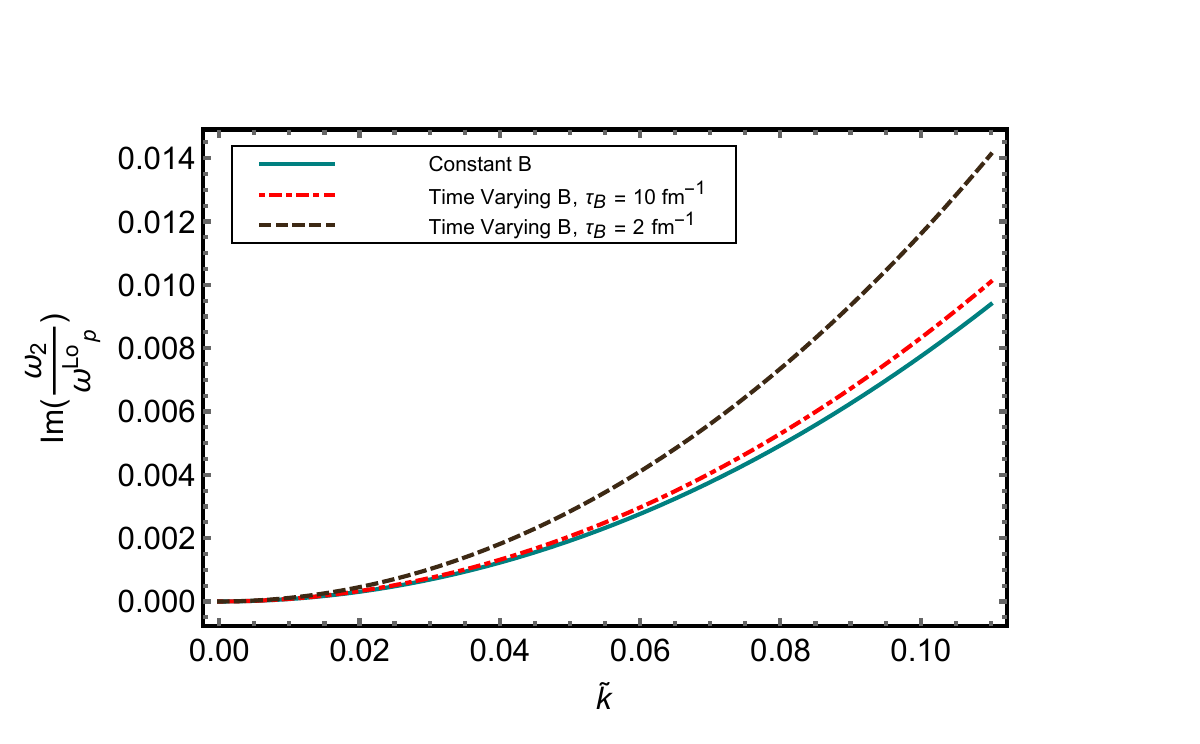}
    \hspace{-2.5cm}
    \caption{\small Imaginary modes of $\omega_1$ (Left Panel) and $\omega_2$ (Right Panel) with different conditions of the magnetic field, Constant magnetic field, time varying magnetic field with $\tau_b = 10$ $fm^{-1}$ and $\tau_b = 2$ $fm^{-1}$. The graph is plotted at $T = 2T_C$ with $T_C = 0.17$ GeV.}
\label{f2}
\end{figure*}

\begin{figure}
    \includegraphics[width=0.571\textwidth]{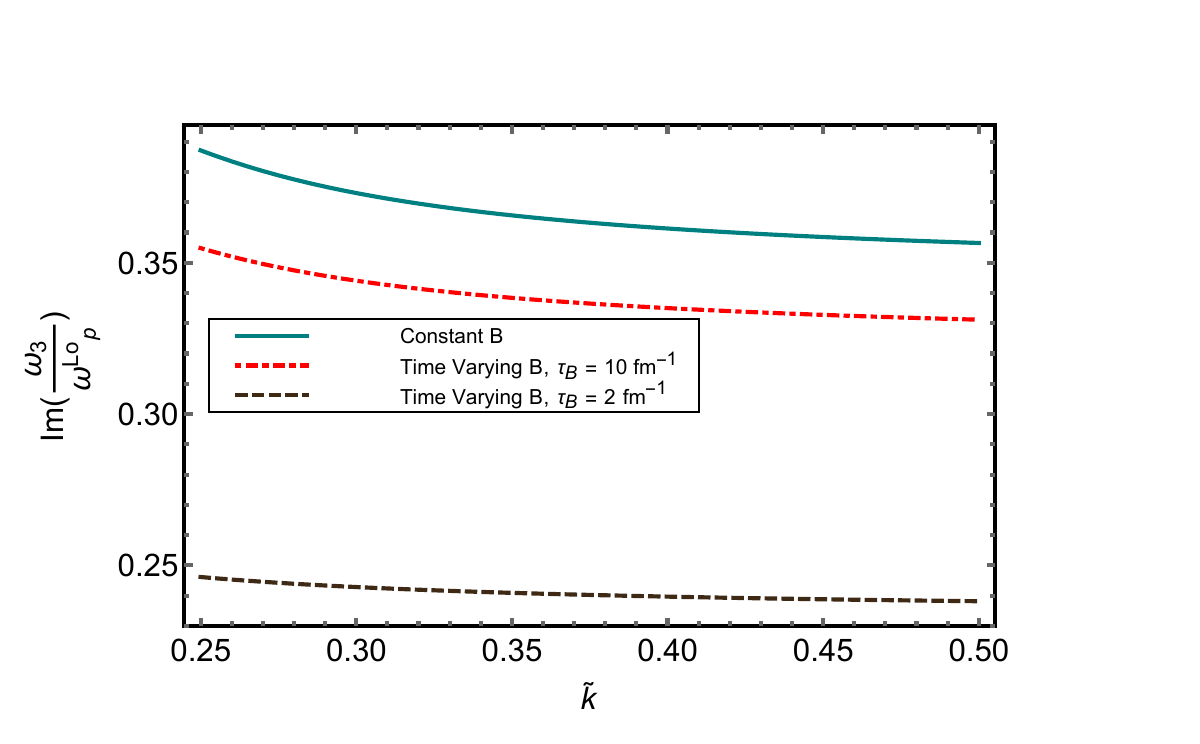}
    \caption{\small Imaginary mode of $\omega_3$ with different conditions of the magnetic field, Constant magnetic field, time varying magnetic field with $\tau_b = 10$ $fm^{-1}$ and $\tau_b = 2$ $fm^{-1}$. The graph is plotted at $T = 2T_C$ with $T_C = 0.17$ GeV.}
\label{f3}
\end{figure}

The structure function corresponding to $P^{ij}_b$ and along the direction of the magnetic fields is shown to be zero, $\gamma_0 =0$ and the rest, $\alpha_0, \beta_0$ and $\delta_0$ are shown to be related to the conductivities.
\subsection{Propagator and the dispersion relation}
The inverse of the propagator can be obtained from the relation with the Polarization tensor.
\begin{align}\label{1.21}
    [\Delta^{-1}]^{ij} = \alpha_1 P^{ij}_L +\beta_1 P^{ij}_T +\delta_1 P^{ij}_{bk},
\end{align}
where the coefficients are obtained as, 
\begin{align}\label{1.22}
    &\alpha_1 = (-\omega^2 +\alpha_0), &&\beta_1 = [(k^2 -\omega^2) + \beta_0], &&&\delta_1 = \delta_0,    
\end{align}
here the coefficient $\gamma_1$ corresponding to $P^{ij}_b$ is $\gamma_1 =\gamma_0 =0$.

The tensor exists in the same space as its inverse; hence, we can expand the propagator along the same components and obtain, 
\begin{align}\label{1.23}
    [\Delta]^{ij} = \alpha_2 P^{ij}_L +\beta_2 P^{ij}_T +\delta_2 P^{ij}_{bk},
\end{align}
using the property, $[\Delta^{-1}]^{ij} [\Delta]_{j\eta} = \delta^{i}_{\eta}$ and using the following contractions, 
\begin{align}\label{1.24}
    &(P_T)^{ij} (P_T)_{j\eta} = [\delta^i_{\eta} -\frac{k^i k_{\eta}}{k^2}],\\ &(P_L)^{ij} (P_L)_{j\eta} = \frac{k^ik_{\eta}}{k^2},\\
    &(P_T)^{ij} (P_{bk})_{j\eta} = \epsilon^i_{\eta \theta}b^{\theta} -\frac{k^i k^j}{k^2} \epsilon_{j\eta \theta} b^{\theta},\\ &(P_T)^{ij} (P_L)_{j\eta} =0,\\ 
    &(P_L)^{ij} (P_{bk})_{j\eta} = \frac{k^ik^j}{k^2} \epsilon_{j\eta \theta}b^{\theta},\\ &(P_{bk})^{ij} (P_{bk})_{j\eta} = -\delta^i_{\eta} +b_{\eta} b^{i},
\end{align}

\begin{figure*}
    \centering
    \hspace{-2.5cm}
    \includegraphics[width=0.571\textwidth]{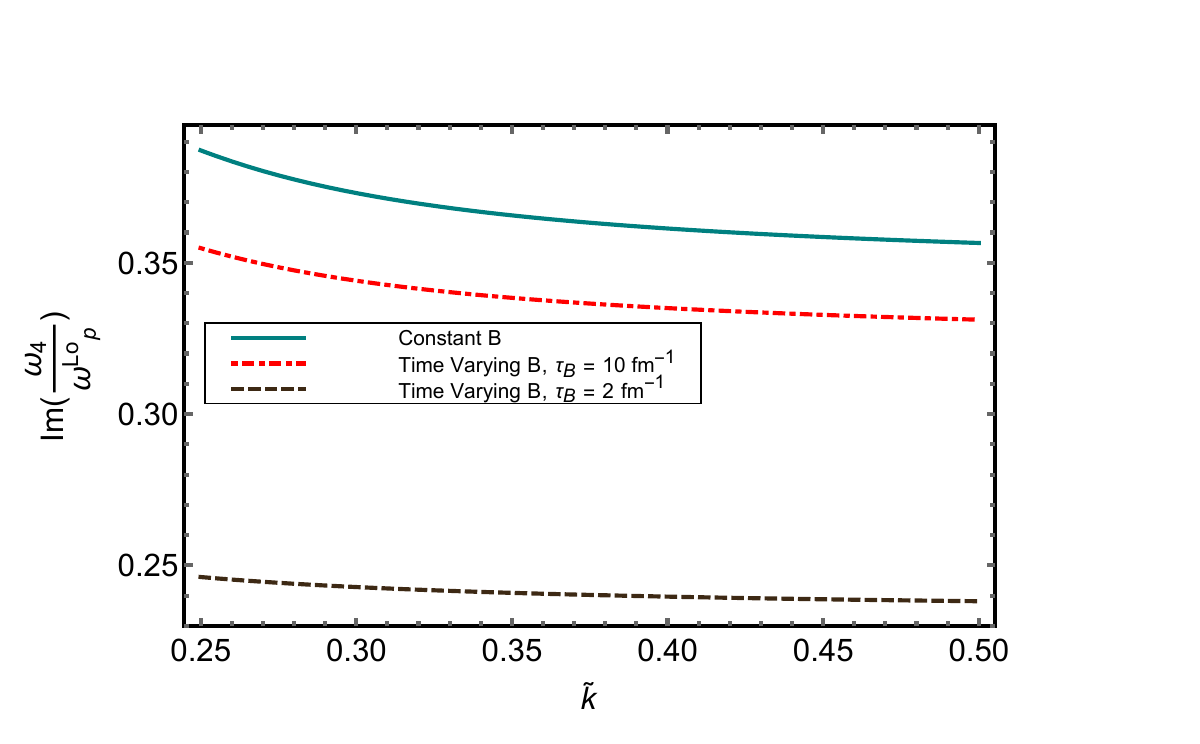}
    \hspace{-.5cm}
    \includegraphics[width=0.571\textwidth]{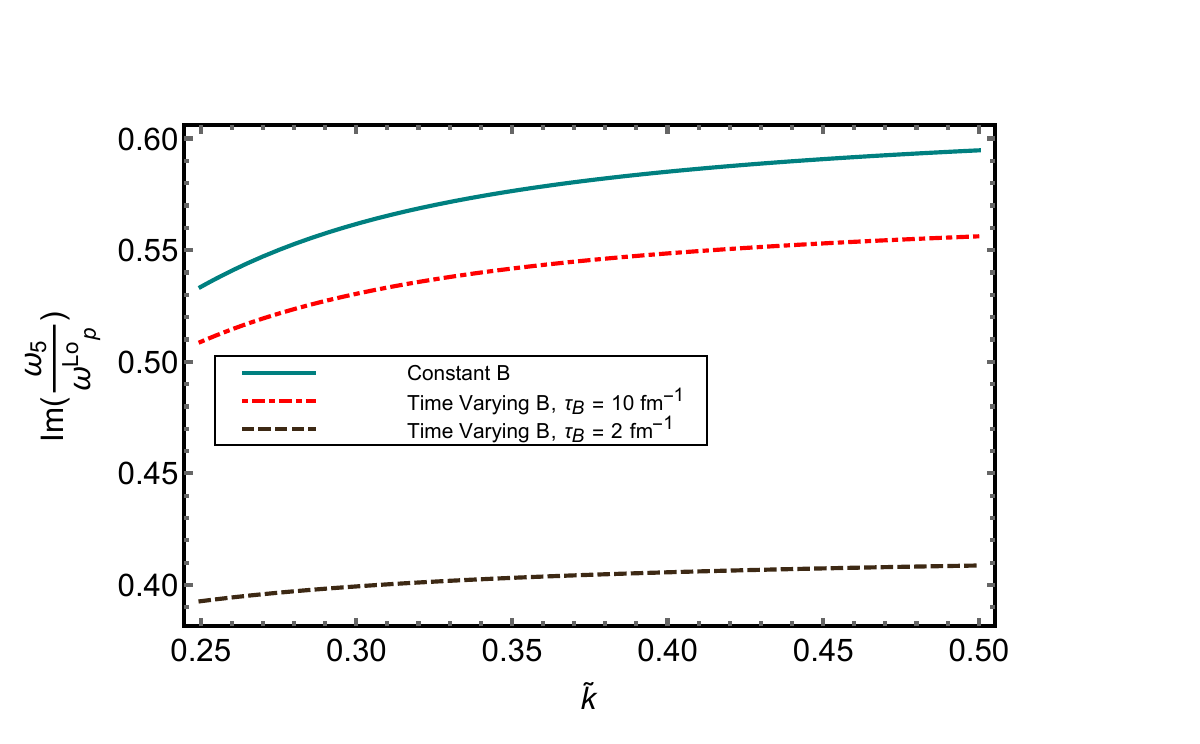}
    \hspace{-2.5cm}
    \caption{\small Imaginary modes of $\omega_4$ (Left Panel) and $\omega_5$ (Right Panel) with different conditions of the magnetic field, Constant magnetic field, time varying magnetic field with $\tau_b = 10$ $fm^{-1}$ and $\tau_b = 2$ $fm^{-1}$. The graph is plotted at $T = 2T_C$ with $T_C = 0.17$ GeV. }
\label{f4}
\end{figure*}

we obtain the coefficients of the propagator as, 
\begin{align}\label{1.25}
    &\alpha_2 = -\frac{-2\alpha_1\beta_1-2\beta_1^2-\delta_12}{(\alpha_1 +\beta_1)(\alpha_1 \beta_1 +\delta_1^2)},\\
    &\beta_2 = -\frac{-\alpha_1^2 -\alpha_1 \beta_1 +\delta_1^2}{(\alpha_1 +\beta_1)(\alpha_1 \beta_1 +\delta_1^2)},\\
    &\delta_2 = -\frac{(\alpha_1 +\beta_1)\delta_1}{(\alpha_1 +\beta_1)(\alpha_1 \beta_1 +\delta_1^2)}.
\end{align}
To find the poles and the dispersion relations, we can solve for the denominator to be zero, which are factorized as,
\begin{equation}\label{1.26}
    \Delta_1 = \alpha_1 +\beta_1 =0 ; \Delta_2 =\alpha_1 \beta_1 +\delta_1^2.
\end{equation} 
The dispersion relations of $\Delta_1$ are found by setting $(-2\omega^2 +k^2+2i\omega \sigma_e)=0$,
\begin{align}\label{1.27}
    &\omega_1 = \frac{i\sigma_e + i\sqrt{(\sigma_e^2 -2k^2)}}{2},\\ \nonumber
    &\omega_2 = \frac{i\sigma_e - i\sqrt{(\sigma_e^2 -2k^2)}}{2}.
\end{align}
The dispersion relations of $\Delta_2$ can be found by setting $(-\omega^2 +i\omega \sigma_e)(k^2-\omega^2 +i\omega \sigma_e)-\omega^2 \sigma_H^2=0$ and solving for $\omega$,
\begin{align}\label{1.28}
        &\omega_3 = \frac{2i\omega \sigma_e}{3}-\frac{2^{1/3} [(i\omega \sigma_e)^2-3(i\omega \sigma_H)^2-3k^2]}{3l}\\ \nonumber &+\frac{l}{3\times 2^{1/3}},\\ \nonumber 
        &\omega_4 = \frac{2i\omega \sigma_e}{3}+\frac{(1+i\sqrt{3}) [(i\omega \sigma_e)^2-3(i\omega \sigma_H)^2-3k^2]}{2^{2/3} \times 3l}\\ \nonumber &-\frac{(1-i\sqrt{3})l}{6\times 2^{1/3}},\\ 
        &\omega_5 = \frac{2i\omega \sigma_e}{3}+\frac{(1-i\sqrt{3}) [(i\omega \sigma_e)^2-3(i\omega \sigma_H)^2-3k^2]}{2^{2/3} \times 3l}\\ \nonumber &-\frac{(1+i\sqrt{3})l}{6\times 2^{1/3}},\nonumber
\end{align}
where, 
    \begin{align}\label{1.29}
        l=&\Big[2i(i\omega \sigma_e)^3+18i(( i\omega \sigma_e) (i \omega \sigma^2_h)^2)-9i(i\omega \sigma_e)k^2+\\ \nonumber &\big(4[(i\omega \sigma_e)^2-3(i\omega \sigma_H)^2-3k^2]^3+2i(i\omega \sigma_e)^3\\ \nonumber &+18i(i\omega \sigma_e)(i\omega \sigma_H)^2-9i(i\omega \sigma_e)k^2]^2\big)^{1/2} \Big]^{1/3}.
    \end{align}
With $\omega_6=0$. The hot QCD medium in the presence of a weak magnetic field shows five modes; among these, only the positive energy dispersion relations are considered. The negative modes can be shown to arise due to the ambiguity in the positive and negative directions of the orthonormal basis we have constructed.

\section{Energy Loss of a heavy quark}

\begin{figure*}
    \centering
    \hspace{-2.5cm}
    \includegraphics[width=0.571\textwidth]{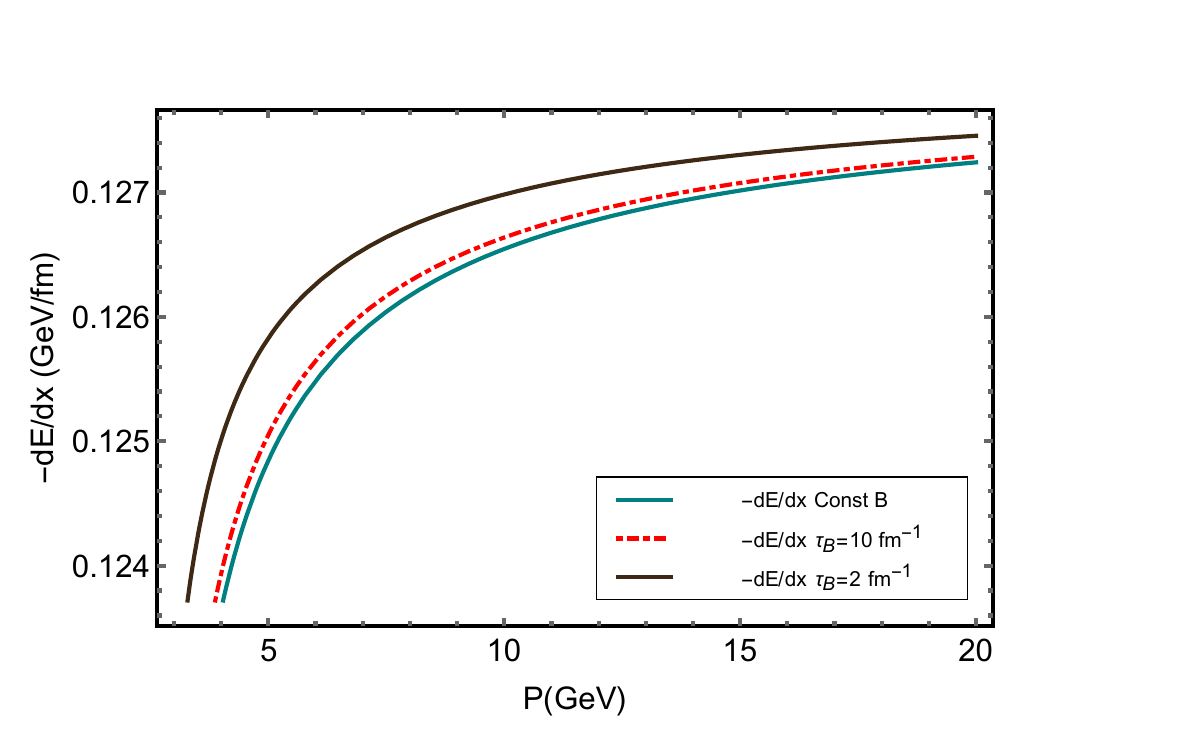}
    \hspace{-.5cm}
    \includegraphics[width=0.571\textwidth]{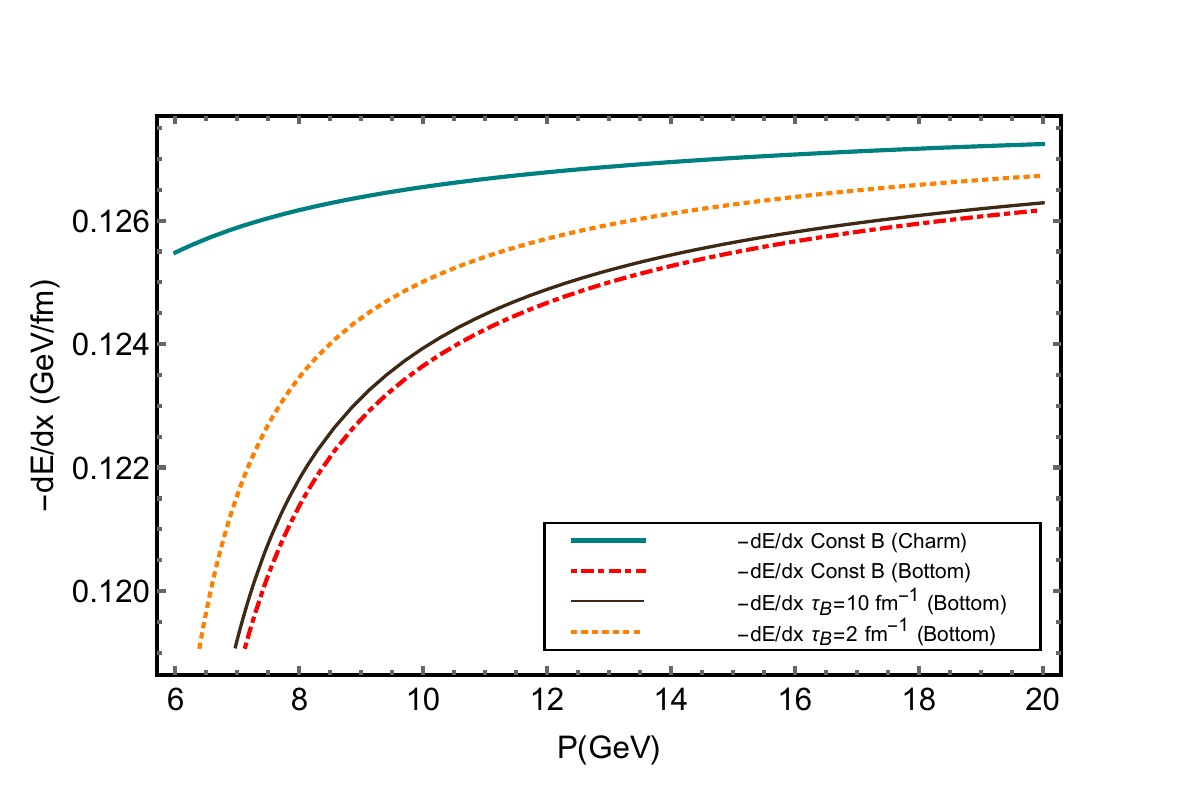}
    \hspace{-2.5cm}
    \caption{\small Energy loss of Charm quark (Left panel) and Bottom Quark (Right Panel), for different conditions of the magnetic field, Constant magnetic field, time varying magnetic field with $\tau_b =10$ $fm^{-1}$ and with $\tau_b = 2$ $fm^{-1}$. The graph is plotted at $T = 2T_C$ with $T_C = 0.17$ GeV. }
\label{f5}
\end{figure*}

A heavy quark moving in the medium loses its energy and is described by Wong's equation. The classical equations which describe the evolution of a point charged particle are given by, 
\begin{align}\label{1.30}
    &\frac{dx^{\mu}(\tau)}{d\tau}=u^{\mu}(\tau),\\ \nonumber &\frac{dp^{\mu}(\tau)}{d\tau}=gq^a(\tau)F_a^{\mu \nu}(x(\tau))u_{\nu}(\tau),\\ \nonumber
    &\frac{dq^a(\tau)}{d\tau} = -gf^{abc} u_{\mu}(\tau)A_b^{\mu}(x(\tau))q_c(\tau),
\end{align}
where $\tau,$ $x(\tau),$ $p^{\mu}(\tau)$ are the proper time, the trajectory and the four momentum of the heavy patron, respectively. The four velocity, $u^{\mu}$ is given by, $u^{\mu} = \frac{p^{\mu}(\tau)}{m}$, with $m$ being the mass of the heavy particle. 

The energy loss expression of the heavy patrons is obtained by imposing two conditions on Wong's equation. First, the gauge condition that the $u^{\mu}A^{\mu}_a = 0$ implies that the third Wong's equation goes to zero, or the charge is independent of the proper time and a constant. The second condition is that the quark's momentum and energy evolve in time without changing the magnitude of its velocity while interacting with the field. Now considering the second Wong's equation, it simplifies under these conditions, 
\begin{align}\label{1.31}
    -\frac{dE}{dX} = gq^a \frac{\textbf{u}}{|\textbf{u}|} .\textbf{E}^a_{ind}(X).
\end{align}
The induced field, $E_{ind}(X)$, can be obtained from the equation of the propagator, 
\begin{align}\label{1.32}
    E_{ind}^j (K) = i\omega[\Delta^{ij}(k)] J_i^{ext}(K).
\end{align}
The external current of a point charge in Fourier space is given by,
\begin{align}\label{1.33}
    \textbf{J}_{ext} (K) = \frac{igq^a \textbf{u}}{\omega-\textbf{u}.\textbf{k}+i0^{+}}.
\end{align}

Contracting the external current with the propagator, we obtain the induced electric field,
\begin{align}\label{1.34}
    &E_{ind}^j (X) = -igq\int d\omega d^3\textbf{k} (\alpha_2 P^{ij}_L u_i\\ \nonumber &+\beta_2 P^{ij}_T u_i+\delta_2 P^{ij}_{bk} u_i) \frac{\exp{i(k^ix_i-\omega t)}}{\omega-u^i.k_i+i0^{+}},
\end{align}
using this, the energy loss of the heavy patron can be written as, 
\begin{align}\label{1.35}
    &-\frac{dE}{dX} = \frac{C_F \alpha_s}{2\pi^2 |\textbf{u}|} \int d\omega d^3\textbf{k} (\alpha_2 P^{ij}_L u_i u_j 
    +\beta_2 P^{ij}_T u_i u_j\\ &+\delta_2 P^{ij}_{bk} u_i u_j)\frac{\exp{i(k^ix_i-\omega t)}}{\omega-u^i.k_i+i0^{+}},\nonumber
\end{align}
with $C_F =4/3$ being the Casamir invariant and $\alpha_s$ is the QCD coupling constant. It can be seen that the antisymmetric projection tensor does not contribute to the energy loss of heavy quarks due to the projection of $u^i u^j$ being zero, $(P^{ij}_{bk})u^i u^j =0$.

\section{Results and Discussions}\label{V}
The dynamics of the hot QCD matter can be understood through the dispersion relations and can be obtained by solving for $\omega$ in $\Delta_1$ and $\Delta_2$ as given in Eq.~(\ref{1.26}). The solution to $\Delta_1$ has one positive real dispersion relation and two imaginary solutions. The dispersion relations of $\Delta_1$ are related to $\sigma_e$, the Ohmic conductivity. We have plotted the positive real dispersion relation of $\Delta_1$ in the left panel of Fig.~\ref{f1}. The dispersion relation has been normalized with the plasma frequency, $\omega_p^{Lo} = m^{Lo}_D/\sqrt{3}$ and plotted against $\Tilde{k} = k/\omega_p^{Lo}$. The time dependent nature of the magnetic field has been studied by considering three different cases of the magnetic field, constant magnetic field, time dependent magnetic field with two decay rates of $\tau_b = 10,2$ $fm^{-1}$. We see that the effects of the magnetic field are more pronounced in the low $\Tilde{k}$ region. The time dependent nature of the magnetic field is seen to increase the dispersion relation with a faster decaying magnetic field, $\tau_b =2$ $fm^{-1}$ having the larger effect. The dispersion relation of $\Delta_2$ has one positive real part, $\omega_3$, that has been plotted in the right panel of Fig.~\ref{f1}. The dispersion relation of $\Delta_2$ is related to both the Ohmic and Hall conductivity, with the Hall one related through $\delta_1=i\omega \sigma_H$. The behaviour of $\omega_3$ is similar to that of $\omega_2$ while significantly larger than $\omega_2$.  

The imaginary dispersion relations are plotted in Figs.~\ref{f2}-~\ref{f4}. The imaginary part of $\omega_1$ is plotted at a constant temperature of $t=2T_c$ in the left panel of Fig.~\ref{f2}. The time dependent nature of the magnetic field is seen by looking at the three cases of the magnetic field, similar to the analysis in Fig.~\ref{f1}. It is seen to decrease with the normalized wavenumber, $\Tilde{k}$. The time dependent nature of the magnetic field decreases the dispersion relation with a faster decaying magnetic field, $\tau_b=10$ $fm^{-1}$, decreasing the dispersion relation the most. In the right panel of Fig.~\ref{f2}, we have plotted the other imaginary dispersion relation in $\omega_2$. The imaginary part of $\omega_2$ increases with $\Tilde{k}$, with the faster decaying magnetic field having the most impact on the dispersion relation. The growing positive imaginary mode of $\omega_2$ may lead to instabilities in the medium. 

The dispersion relations of $\Delta_2$ consist of three complex solutions. The imaginary part of $\omega_3$ and $\omega_4$ are plotted in Fig.~\ref{f3} and the left panel of Fig.~\ref{f4}, respectively. These results underscore the influence of the time-dependent magnetic field, indicating that the decay rate of the magnetic field significantly impacts the dispersion relation. In the right panel of Fig.~\ref{f4}, we have plotted the imaginary part of  $\omega_5$. It is seen to be growing with $\Tilde{k}$ and may perhaps lead to instabilities in the medium.

The energy loss of heavy quarks is plotted in Fig.~\ref{f5} for the Charm (Left panel) and Bottom quarks (Right panel). The introduction of time varying magnetic field increases the energy loss, with a fast decaying magnetic field having the most significant effect. The difference between the energy loss of Charm and Bottom quark is also observed and found to be in accordance with an earlier work~\cite{YousufJamal:2019pen}.
The lighter of the two quarks, the Charm quark, experiences greater energy loss within the same momentum range when compared to its heavier counterpart, the Bottom quark. This arises due to the fact that a quark with a greater mass travels at a slower velocity when the momentum remains constant.
Consequently, a particle moving slowly interacts less with the medium and experiences less energy loss.

\section{Conclusion and Outlook}\label{VI}
The collective modes of a hot QCD medium are obtained through the polarization tensor in a background weak time dependent magnetic field within a linear transport theory. The dispersion relation and the collective modes are obtained by invoking Maxwell's equation. The hot QCD medium interactions are incorporated through the collision term in the relaxation time approximation. The dependence of the collective modes on the wavenumber is plotted, and the effects of the magnetic field are prominent in the lower wavenumber regions. The imaginary parts of the collective modes are shown to be significantly affected by the magnetic field and are also dependent on the temperature of the medium through the conductivities. We have also observed the energy loss of the heavy quarks moving through a medium in background time dependent electromagnetic fields. The decay rate of the magnetic field is seen to impact the energy loss of the heavy quarks significantly. We observe that the heavier quark (Bottom) loses less energy than the lighter (Charm) quark, the heavier particle at a fixed value of momentum $p$ travels at a lower velocity, and a slowly moving particle loses less energy in the medium. These findings are consistent with existing works with constant magnetic fields.

An immediate extension of the work would be to look at the effects of the equation of state on the collective behaviour. The effects of momentum anisotropy on the system would also be a good extension of the project. Working with more realistic collision kernels such as the BGK and modified BGK would be another direction where our future investigations will focus. The realization of a hot QCD medium in terms of a refractive index would also be taken up in future. Furthermore, investigating $R_{AA}$ in future work could be crucial as it provides a vital link between theoretical predictions and experimental findings.
\newline
\section*{Acknowledgments}
GKK acknowledges the Indian Institute of Technology, Gandhinagar (IIT GN), for the Overseas Research Experience Fellowship to visit the Institute of Nuclear Physics PAN and the hospitality of INP PAN, where a portion of the work was completed. V. C. acknowledges the SERB Core Research Grant (CRG) [CRG/2020/002320]. 
\bibliography{ref}{}

\begin{thebibliography}{45}%
\makeatletter
\providecommand \@ifxundefined [1]{%
 \@ifx{#1\undefined}
}%
\providecommand \@ifnum [1]{%
 \ifnum #1\expandafter \@firstoftwo
 \else \expandafter \@secondoftwo
 \fi
}%
\providecommand \@ifx [1]{%
 \ifx #1\expandafter \@firstoftwo
 \else \expandafter \@secondoftwo
 \fi
}%
\providecommand \natexlab [1]{#1}%
\providecommand \enquote  [1]{``#1''}%
\providecommand \bibnamefont  [1]{#1}%
\providecommand \bibfnamefont [1]{#1}%
\providecommand \citenamefont [1]{#1}%
\providecommand \href@noop [0]{\@secondoftwo}%
\providecommand \href [0]{\begingroup \@sanitize@url \@href}%
\providecommand \@href[1]{\@@startlink{#1}\@@href}%
\providecommand \@@href[1]{\endgroup#1\@@endlink}%
\providecommand \@sanitize@url [0]{\catcode `\\12\catcode `\$12\catcode `\&12\catcode `\#12\catcode `\^12\catcode `\_12\catcode `\%12\relax}%
\providecommand \@@startlink[1]{}%
\providecommand \@@endlink[0]{}%
\providecommand \url  [0]{\begingroup\@sanitize@url \@url }%
\providecommand \@url [1]{\endgroup\@href {#1}{\urlprefix }}%
\providecommand \urlprefix  [0]{URL }%
\providecommand \Eprint [0]{\href }%
\providecommand \doibase [0]{http://dx.doi.org/}%
\providecommand \selectlanguage [0]{\@gobble}%
\providecommand \bibinfo  [0]{\@secondoftwo}%
\providecommand \bibfield  [0]{\@secondoftwo}%
\providecommand \translation [1]{[#1]}%
\providecommand \BibitemOpen [0]{}%
\providecommand \bibitemStop [0]{}%
\providecommand \bibitemNoStop [0]{.\EOS\space}%
\providecommand \EOS [0]{\spacefactor3000\relax}%
\providecommand \BibitemShut  [1]{\csname bibitem#1\endcsname}%
\let\auto@bib@innerbib\@empty
\bibitem [{\citenamefont {Adams}\ \emph {et~al.}(2005)\citenamefont {Adams} \emph {et~al.}}]{Adams:2005dq}%
  \BibitemOpen
  \bibfield  {author} {\bibinfo {author} {\bibfnamefont {J.}~\bibnamefont {Adams}} \emph {et~al.} (\bibinfo {collaboration} {STAR}),\ }\href {\doibase 10.1016/j.nuclphysa.2005.03.085} {\bibfield  {journal} {\bibinfo  {journal} {Nucl. Phys. A}\ }\textbf {\bibinfo {volume} {757}},\ \bibinfo {pages} {102} (\bibinfo {year} {2005})},\ \Eprint {http://arxiv.org/abs/nucl-ex/0501009} {arXiv:nucl-ex/0501009} \BibitemShut {NoStop}%
\bibitem [{\citenamefont {Ryu}\ \emph {et~al.}(2015)\citenamefont {Ryu}, \citenamefont {Paquet}, \citenamefont {Shen}, \citenamefont {Denicol}, \citenamefont {Schenke}, \citenamefont {Jeon},\ and\ \citenamefont {Gale}}]{Ryu:2015vwa}%
  \BibitemOpen
  \bibfield  {author} {\bibinfo {author} {\bibfnamefont {S.}~\bibnamefont {Ryu}}, \bibinfo {author} {\bibfnamefont {J.~F.}\ \bibnamefont {Paquet}}, \bibinfo {author} {\bibfnamefont {C.}~\bibnamefont {Shen}}, \bibinfo {author} {\bibfnamefont {G.~S.}\ \bibnamefont {Denicol}}, \bibinfo {author} {\bibfnamefont {B.}~\bibnamefont {Schenke}}, \bibinfo {author} {\bibfnamefont {S.}~\bibnamefont {Jeon}}, \ and\ \bibinfo {author} {\bibfnamefont {C.}~\bibnamefont {Gale}},\ }\href {\doibase 10.1103/PhysRevLett.115.132301} {\bibfield  {journal} {\bibinfo  {journal} {Phys. Rev. Lett.}\ }\textbf {\bibinfo {volume} {115}},\ \bibinfo {pages} {132301} (\bibinfo {year} {2015})},\ \Eprint {http://arxiv.org/abs/1502.01675} {arXiv:1502.01675 [nucl-th]} \BibitemShut {NoStop}%
\bibitem [{\citenamefont {Denicol}\ \emph {et~al.}(2016)\citenamefont {Denicol}, \citenamefont {Monnai},\ and\ \citenamefont {Schenke}}]{Denicol:2015nhu}%
  \BibitemOpen
  \bibfield  {author} {\bibinfo {author} {\bibfnamefont {G.}~\bibnamefont {Denicol}}, \bibinfo {author} {\bibfnamefont {A.}~\bibnamefont {Monnai}}, \ and\ \bibinfo {author} {\bibfnamefont {B.}~\bibnamefont {Schenke}},\ }\href {\doibase 10.1103/PhysRevLett.116.212301} {\bibfield  {journal} {\bibinfo  {journal} {Phys. Rev. Lett.}\ }\textbf {\bibinfo {volume} {116}},\ \bibinfo {pages} {212301} (\bibinfo {year} {2016})},\ \Eprint {http://arxiv.org/abs/1512.01538} {arXiv:1512.01538 [nucl-th]} \BibitemShut {NoStop}%
\bibitem [{\citenamefont {Acharya}\ \emph {et~al.}(2020)\citenamefont {Acharya} \emph {et~al.}}]{Acharya:2019ijj}%
  \BibitemOpen
  \bibfield  {author} {\bibinfo {author} {\bibfnamefont {S.}~\bibnamefont {Acharya}} \emph {et~al.} (\bibinfo {collaboration} {ALICE}),\ }\href {\doibase 10.1103/PhysRevLett.125.022301} {\bibfield  {journal} {\bibinfo  {journal} {Phys. Rev. Lett.}\ }\textbf {\bibinfo {volume} {125}},\ \bibinfo {pages} {022301} (\bibinfo {year} {2020})},\ \Eprint {http://arxiv.org/abs/1910.14406} {arXiv:1910.14406 [nucl-ex]} \BibitemShut {NoStop}%
\bibitem [{\citenamefont {Adam}\ \emph {et~al.}(2019)\citenamefont {Adam} \emph {et~al.}}]{Adam:2019wnk}%
  \BibitemOpen
  \bibfield  {author} {\bibinfo {author} {\bibfnamefont {J.}~\bibnamefont {Adam}} \emph {et~al.} (\bibinfo {collaboration} {STAR}),\ }\href {\doibase 10.1103/PhysRevLett.123.162301} {\bibfield  {journal} {\bibinfo  {journal} {Phys. Rev. Lett.}\ }\textbf {\bibinfo {volume} {123}},\ \bibinfo {pages} {162301} (\bibinfo {year} {2019})},\ \Eprint {http://arxiv.org/abs/1905.02052} {arXiv:1905.02052 [nucl-ex]} \BibitemShut {NoStop}%
\bibitem [{\citenamefont {Tuchin}(2013)}]{Tuchin:2013apa}%
  \BibitemOpen
  \bibfield  {author} {\bibinfo {author} {\bibfnamefont {K.}~\bibnamefont {Tuchin}},\ }\href {\doibase 10.1103/PhysRevC.88.024911} {\bibfield  {journal} {\bibinfo  {journal} {Phys. Rev. C}\ }\textbf {\bibinfo {volume} {88}},\ \bibinfo {pages} {024911} (\bibinfo {year} {2013})},\ \Eprint {http://arxiv.org/abs/1305.5806} {arXiv:1305.5806 [hep-ph]} \BibitemShut {NoStop}%
\bibitem [{\citenamefont {McLerran}\ and\ \citenamefont {Skokov}(2014)}]{McLerran:2013hla}%
  \BibitemOpen
  \bibfield  {author} {\bibinfo {author} {\bibfnamefont {L.}~\bibnamefont {McLerran}}\ and\ \bibinfo {author} {\bibfnamefont {V.}~\bibnamefont {Skokov}},\ }\href {\doibase 10.1016/j.nuclphysa.2014.05.008} {\bibfield  {journal} {\bibinfo  {journal} {Nucl. Phys. A}\ }\textbf {\bibinfo {volume} {929}},\ \bibinfo {pages} {184} (\bibinfo {year} {2014})},\ \Eprint {http://arxiv.org/abs/1305.0774} {arXiv:1305.0774 [hep-ph]} \BibitemShut {NoStop}%
\bibitem [{\citenamefont {Stewart}\ and\ \citenamefont {Tuchin}(2021)}]{Stewart:2021mjz}%
  \BibitemOpen
  \bibfield  {author} {\bibinfo {author} {\bibfnamefont {E.}~\bibnamefont {Stewart}}\ and\ \bibinfo {author} {\bibfnamefont {K.}~\bibnamefont {Tuchin}},\ }\href@noop {} {\  (\bibinfo {year} {2021})},\ \Eprint {http://arxiv.org/abs/2106.09124} {arXiv:2106.09124 [nucl-th]} \BibitemShut {NoStop}%
\bibitem [{\citenamefont {Tuchin}(2020)}]{Tuchin:2019gkg}%
  \BibitemOpen
  \bibfield  {author} {\bibinfo {author} {\bibfnamefont {K.}~\bibnamefont {Tuchin}},\ }\href {\doibase 10.1103/PhysRevC.102.014908} {\bibfield  {journal} {\bibinfo  {journal} {Phys. Rev. C}\ }\textbf {\bibinfo {volume} {102}},\ \bibinfo {pages} {014908} (\bibinfo {year} {2020})},\ \Eprint {http://arxiv.org/abs/1911.01357} {arXiv:1911.01357 [hep-ph]} \BibitemShut {NoStop}%
\bibitem [{\citenamefont {Yan}\ and\ \citenamefont {Huang}(2021)}]{Yan:2021zjc}%
  \BibitemOpen
  \bibfield  {author} {\bibinfo {author} {\bibfnamefont {L.}~\bibnamefont {Yan}}\ and\ \bibinfo {author} {\bibfnamefont {X.-G.}\ \bibnamefont {Huang}},\ }\href@noop {} {\  (\bibinfo {year} {2021})},\ \Eprint {http://arxiv.org/abs/2104.00831} {arXiv:2104.00831 [nucl-th]} \BibitemShut {NoStop}%
\bibitem [{\citenamefont {Feng}(2017)}]{Feng:2017tsh}%
  \BibitemOpen
  \bibfield  {author} {\bibinfo {author} {\bibfnamefont {B.}~\bibnamefont {Feng}},\ }\href {\doibase 10.1103/PhysRevD.96.036009} {\bibfield  {journal} {\bibinfo  {journal} {Phys. Rev. D}\ }\textbf {\bibinfo {volume} {96}},\ \bibinfo {pages} {036009} (\bibinfo {year} {2017})}\BibitemShut {NoStop}%
\bibitem [{\citenamefont {Thakur}\ and\ \citenamefont {Srivastava}(2019)}]{Thakur:2019bnf}%
  \BibitemOpen
  \bibfield  {author} {\bibinfo {author} {\bibfnamefont {L.}~\bibnamefont {Thakur}}\ and\ \bibinfo {author} {\bibfnamefont {P.}~\bibnamefont {Srivastava}},\ }\href {\doibase 10.1103/PhysRevD.100.076016} {\bibfield  {journal} {\bibinfo  {journal} {Phys. Rev. D}\ }\textbf {\bibinfo {volume} {100}},\ \bibinfo {pages} {076016} (\bibinfo {year} {2019})},\ \Eprint {http://arxiv.org/abs/1910.12087} {arXiv:1910.12087 [hep-ph]} \BibitemShut {NoStop}%
\bibitem [{\citenamefont {Dey}\ \emph {et~al.}(2019)\citenamefont {Dey}, \citenamefont {Satapathy}, \citenamefont {Murmu},\ and\ \citenamefont {Ghosh}}]{Dey:2019axu}%
  \BibitemOpen
  \bibfield  {author} {\bibinfo {author} {\bibfnamefont {J.}~\bibnamefont {Dey}}, \bibinfo {author} {\bibfnamefont {S.}~\bibnamefont {Satapathy}}, \bibinfo {author} {\bibfnamefont {P.}~\bibnamefont {Murmu}}, \ and\ \bibinfo {author} {\bibfnamefont {S.}~\bibnamefont {Ghosh}},\ }\href@noop {} {\  (\bibinfo {year} {2019})},\ \Eprint {http://arxiv.org/abs/1907.11164} {arXiv:1907.11164 [hep-ph]} \BibitemShut {NoStop}%
\bibitem [{\citenamefont {Hattori}\ \emph {et~al.}(2017)\citenamefont {Hattori}, \citenamefont {Li}, \citenamefont {Satow},\ and\ \citenamefont {Yee}}]{Hattori:2016lqx}%
  \BibitemOpen
  \bibfield  {author} {\bibinfo {author} {\bibfnamefont {K.}~\bibnamefont {Hattori}}, \bibinfo {author} {\bibfnamefont {S.}~\bibnamefont {Li}}, \bibinfo {author} {\bibfnamefont {D.}~\bibnamefont {Satow}}, \ and\ \bibinfo {author} {\bibfnamefont {H.-U.}\ \bibnamefont {Yee}},\ }\href {\doibase 10.1103/PhysRevD.95.076008} {\bibfield  {journal} {\bibinfo  {journal} {Phys. Rev. D}\ }\textbf {\bibinfo {volume} {95}},\ \bibinfo {pages} {076008} (\bibinfo {year} {2017})},\ \Eprint {http://arxiv.org/abs/1610.06839} {arXiv:1610.06839 [hep-ph]} \BibitemShut {NoStop}%
\bibitem [{\citenamefont {Fukushima}\ and\ \citenamefont {Hidaka}(2018)}]{Fukushima:2017lvb}%
  \BibitemOpen
  \bibfield  {author} {\bibinfo {author} {\bibfnamefont {K.}~\bibnamefont {Fukushima}}\ and\ \bibinfo {author} {\bibfnamefont {Y.}~\bibnamefont {Hidaka}},\ }\href {\doibase 10.1103/PhysRevLett.120.162301} {\bibfield  {journal} {\bibinfo  {journal} {Phys. Rev. Lett.}\ }\textbf {\bibinfo {volume} {120}},\ \bibinfo {pages} {162301} (\bibinfo {year} {2018})},\ \Eprint {http://arxiv.org/abs/1711.01472} {arXiv:1711.01472 [hep-ph]} \BibitemShut {NoStop}%
\bibitem [{\citenamefont {Dash}\ \emph {et~al.}(2020)\citenamefont {Dash}, \citenamefont {Samanta}, \citenamefont {Dey}, \citenamefont {Gangopadhyaya}, \citenamefont {Ghosh},\ and\ \citenamefont {Roy}}]{Dash:2020vxk}%
  \BibitemOpen
  \bibfield  {author} {\bibinfo {author} {\bibfnamefont {A.}~\bibnamefont {Dash}}, \bibinfo {author} {\bibfnamefont {S.}~\bibnamefont {Samanta}}, \bibinfo {author} {\bibfnamefont {J.}~\bibnamefont {Dey}}, \bibinfo {author} {\bibfnamefont {U.}~\bibnamefont {Gangopadhyaya}}, \bibinfo {author} {\bibfnamefont {S.}~\bibnamefont {Ghosh}}, \ and\ \bibinfo {author} {\bibfnamefont {V.}~\bibnamefont {Roy}},\ }\href {\doibase 10.1103/PhysRevD.102.016016} {\bibfield  {journal} {\bibinfo  {journal} {Phys. Rev. D}\ }\textbf {\bibinfo {volume} {102}},\ \bibinfo {pages} {016016} (\bibinfo {year} {2020})},\ \Eprint {http://arxiv.org/abs/2002.08781} {arXiv:2002.08781 [nucl-th]} \BibitemShut {NoStop}%
\bibitem [{\citenamefont {Kurian}\ and\ \citenamefont {Chandra}(2017)}]{Kurian:2017yxj}%
  \BibitemOpen
  \bibfield  {author} {\bibinfo {author} {\bibfnamefont {M.}~\bibnamefont {Kurian}}\ and\ \bibinfo {author} {\bibfnamefont {V.}~\bibnamefont {Chandra}},\ }\href {\doibase 10.1103/PhysRevD.96.114026} {\bibfield  {journal} {\bibinfo  {journal} {Phys. Rev. D}\ }\textbf {\bibinfo {volume} {96}},\ \bibinfo {pages} {114026} (\bibinfo {year} {2017})},\ \Eprint {http://arxiv.org/abs/1709.08320} {arXiv:1709.08320 [nucl-th]} \BibitemShut {NoStop}%
\bibitem [{\citenamefont {Kurian}\ and\ \citenamefont {Chandra}(2019)}]{Kurian:2019fty}%
  \BibitemOpen
  \bibfield  {author} {\bibinfo {author} {\bibfnamefont {M.}~\bibnamefont {Kurian}}\ and\ \bibinfo {author} {\bibfnamefont {V.}~\bibnamefont {Chandra}},\ }\href {\doibase 10.1103/PhysRevD.99.116018} {\bibfield  {journal} {\bibinfo  {journal} {Phys. Rev. D}\ }\textbf {\bibinfo {volume} {99}},\ \bibinfo {pages} {116018} (\bibinfo {year} {2019})},\ \Eprint {http://arxiv.org/abs/1902.09200} {arXiv:1902.09200 [nucl-th]} \BibitemShut {NoStop}%
\bibitem [{\citenamefont {Ghosh}\ \emph {et~al.}(2019)\citenamefont {Ghosh}, \citenamefont {Bandyopadhyay}, \citenamefont {Farias}, \citenamefont {Dey},\ and\ \citenamefont {Krein}}]{Ghosh:2019ubc}%
  \BibitemOpen
  \bibfield  {author} {\bibinfo {author} {\bibfnamefont {S.}~\bibnamefont {Ghosh}}, \bibinfo {author} {\bibfnamefont {A.}~\bibnamefont {Bandyopadhyay}}, \bibinfo {author} {\bibfnamefont {R.~L.}\ \bibnamefont {Farias}}, \bibinfo {author} {\bibfnamefont {J.}~\bibnamefont {Dey}}, \ and\ \bibinfo {author} {\bibfnamefont {G.~a.}\ \bibnamefont {Krein}},\ }\href@noop {} {\  (\bibinfo {year} {2019})},\ \Eprint {http://arxiv.org/abs/1911.10005} {arXiv:1911.10005 [hep-ph]} \BibitemShut {NoStop}%
\bibitem [{\citenamefont {Gowthama}\ \emph {et~al.}(2021)\citenamefont {Gowthama}, \citenamefont {Kurian},\ and\ \citenamefont {Chandra}}]{Gowthama:2020ghl}%
  \BibitemOpen
  \bibfield  {author} {\bibinfo {author} {\bibfnamefont {K.~K.}\ \bibnamefont {Gowthama}}, \bibinfo {author} {\bibfnamefont {M.}~\bibnamefont {Kurian}}, \ and\ \bibinfo {author} {\bibfnamefont {V.}~\bibnamefont {Chandra}},\ }\href {\doibase 10.1103/PhysRevD.103.074017} {\bibfield  {journal} {\bibinfo  {journal} {Phys. Rev. D}\ }\textbf {\bibinfo {volume} {103}},\ \bibinfo {pages} {074017} (\bibinfo {year} {2021})},\ \Eprint {http://arxiv.org/abs/2012.07156} {arXiv:2012.07156 [hep-ph]} \BibitemShut {NoStop}%
\bibitem [{\citenamefont {K}\ \emph {et~al.}(2021)\citenamefont {K}, \citenamefont {Kurian},\ and\ \citenamefont {Chandra}}]{K:2021sct}%
  \BibitemOpen
  \bibfield  {author} {\bibinfo {author} {\bibfnamefont {G.~K.}\ \bibnamefont {K}}, \bibinfo {author} {\bibfnamefont {M.}~\bibnamefont {Kurian}}, \ and\ \bibinfo {author} {\bibfnamefont {V.}~\bibnamefont {Chandra}},\ }\href {\doibase 10.1103/PhysRevD.104.094037} {\bibfield  {journal} {\bibinfo  {journal} {Phys. Rev. D}\ }\textbf {\bibinfo {volume} {104}},\ \bibinfo {pages} {094037} (\bibinfo {year} {2021})},\ \Eprint {http://arxiv.org/abs/2108.06791} {arXiv:2108.06791 [hep-ph]} \BibitemShut {NoStop}%
\bibitem [{\citenamefont {Denicol}\ \emph {et~al.}(2014)\citenamefont {Denicol}, \citenamefont {Niemi}, \citenamefont {Bouras}, \citenamefont {Molnar}, \citenamefont {Xu}, \citenamefont {Rischke},\ and\ \citenamefont {Greiner}}]{Denicol:2012vq}%
  \BibitemOpen
  \bibfield  {author} {\bibinfo {author} {\bibfnamefont {G.~S.}\ \bibnamefont {Denicol}}, \bibinfo {author} {\bibfnamefont {H.}~\bibnamefont {Niemi}}, \bibinfo {author} {\bibfnamefont {I.}~\bibnamefont {Bouras}}, \bibinfo {author} {\bibfnamefont {E.}~\bibnamefont {Molnar}}, \bibinfo {author} {\bibfnamefont {Z.}~\bibnamefont {Xu}}, \bibinfo {author} {\bibfnamefont {D.~H.}\ \bibnamefont {Rischke}}, \ and\ \bibinfo {author} {\bibfnamefont {C.}~\bibnamefont {Greiner}},\ }\href {\doibase 10.1103/PhysRevD.89.074005} {\bibfield  {journal} {\bibinfo  {journal} {Phys. Rev. D}\ }\textbf {\bibinfo {volume} {89}},\ \bibinfo {pages} {074005} (\bibinfo {year} {2014})},\ \Eprint {http://arxiv.org/abs/1207.6811} {arXiv:1207.6811 [nucl-th]} \BibitemShut {NoStop}%
\bibitem [{\citenamefont {Kapusta}\ and\ \citenamefont {Torres-Rincon}(2012)}]{Kapusta:2012zb}%
  \BibitemOpen
  \bibfield  {author} {\bibinfo {author} {\bibfnamefont {J.~I.}\ \bibnamefont {Kapusta}}\ and\ \bibinfo {author} {\bibfnamefont {J.~M.}\ \bibnamefont {Torres-Rincon}},\ }\href {\doibase 10.1103/PhysRevC.86.054911} {\bibfield  {journal} {\bibinfo  {journal} {Phys. Rev. C}\ }\textbf {\bibinfo {volume} {86}},\ \bibinfo {pages} {054911} (\bibinfo {year} {2012})},\ \Eprint {http://arxiv.org/abs/1209.0675} {arXiv:1209.0675 [nucl-th]} \BibitemShut {NoStop}%
\bibitem [{\citenamefont {Kurian}(2021)}]{Kurian:2021zyb}%
  \BibitemOpen
  \bibfield  {author} {\bibinfo {author} {\bibfnamefont {M.}~\bibnamefont {Kurian}},\ }\href {\doibase 10.1103/PhysRevD.103.054024} {\bibfield  {journal} {\bibinfo  {journal} {Phys. Rev. D}\ }\textbf {\bibinfo {volume} {103}},\ \bibinfo {pages} {054024} (\bibinfo {year} {2021})},\ \Eprint {http://arxiv.org/abs/2102.00435} {arXiv:2102.00435 [hep-ph]} \BibitemShut {NoStop}%
\bibitem [{\citenamefont {K}\ \emph {et~al.}(2022)\citenamefont {K}, \citenamefont {Kurian},\ and\ \citenamefont {Chandra}}]{K:2022pzc}%
  \BibitemOpen
  \bibfield  {author} {\bibinfo {author} {\bibfnamefont {G.~K.}\ \bibnamefont {K}}, \bibinfo {author} {\bibfnamefont {M.}~\bibnamefont {Kurian}}, \ and\ \bibinfo {author} {\bibfnamefont {V.}~\bibnamefont {Chandra}},\ }\href {\doibase 10.1103/PhysRevD.106.034008} {\bibfield  {journal} {\bibinfo  {journal} {Phys. Rev. D}\ }\textbf {\bibinfo {volume} {106}},\ \bibinfo {pages} {034008} (\bibinfo {year} {2022})},\ \Eprint {http://arxiv.org/abs/2205.14427} {arXiv:2205.14427 [hep-ph]} \BibitemShut {NoStop}%
\bibitem [{\citenamefont {Carrington}\ \emph {et~al.}(2004)\citenamefont {Carrington}, \citenamefont {Fugleberg}, \citenamefont {Pickering},\ and\ \citenamefont {Thoma}}]{Carrington:2003je}%
  \BibitemOpen
  \bibfield  {author} {\bibinfo {author} {\bibfnamefont {M.}~\bibnamefont {Carrington}}, \bibinfo {author} {\bibfnamefont {T.}~\bibnamefont {Fugleberg}}, \bibinfo {author} {\bibfnamefont {D.}~\bibnamefont {Pickering}}, \ and\ \bibinfo {author} {\bibfnamefont {M.}~\bibnamefont {Thoma}},\ }\href {\doibase 10.1139/p04-035} {\bibfield  {journal} {\bibinfo  {journal} {Can. J. Phys.}\ }\textbf {\bibinfo {volume} {82}},\ \bibinfo {pages} {671} (\bibinfo {year} {2004})},\ \Eprint {http://arxiv.org/abs/hep-ph/0312103} {arXiv:hep-ph/0312103} \BibitemShut {NoStop}%
\bibitem [{\citenamefont {Schenke}\ \emph {et~al.}(2006)\citenamefont {Schenke}, \citenamefont {Strickland}, \citenamefont {Greiner},\ and\ \citenamefont {Thoma}}]{Schenke:2006xu}%
  \BibitemOpen
  \bibfield  {author} {\bibinfo {author} {\bibfnamefont {B.}~\bibnamefont {Schenke}}, \bibinfo {author} {\bibfnamefont {M.}~\bibnamefont {Strickland}}, \bibinfo {author} {\bibfnamefont {C.}~\bibnamefont {Greiner}}, \ and\ \bibinfo {author} {\bibfnamefont {M.~H.}\ \bibnamefont {Thoma}},\ }\href {\doibase 10.1103/PhysRevD.73.125004} {\bibfield  {journal} {\bibinfo  {journal} {Phys. Rev. D}\ }\textbf {\bibinfo {volume} {73}},\ \bibinfo {pages} {125004} (\bibinfo {year} {2006})},\ \Eprint {http://arxiv.org/abs/hep-ph/0603029} {arXiv:hep-ph/0603029} \BibitemShut {NoStop}%
\bibitem [{\citenamefont {Carrington}\ \emph {et~al.}(2014)\citenamefont {Carrington}, \citenamefont {Deja},\ and\ \citenamefont {Mrowczynski}}]{Carrington:2014bla}%
  \BibitemOpen
  \bibfield  {author} {\bibinfo {author} {\bibfnamefont {M.~E.}\ \bibnamefont {Carrington}}, \bibinfo {author} {\bibfnamefont {K.}~\bibnamefont {Deja}}, \ and\ \bibinfo {author} {\bibfnamefont {S.}~\bibnamefont {Mrowczynski}},\ }\href {\doibase 10.1103/PhysRevC.90.034913} {\bibfield  {journal} {\bibinfo  {journal} {Phys. Rev. C}\ }\textbf {\bibinfo {volume} {90}},\ \bibinfo {pages} {034913} (\bibinfo {year} {2014})},\ \Eprint {http://arxiv.org/abs/1407.2764} {arXiv:1407.2764 [hep-ph]} \BibitemShut {NoStop}%
\bibitem [{\citenamefont {Romatschke}\ and\ \citenamefont {Strickland}(2003)}]{Romatschke:2003ms}%
  \BibitemOpen
  \bibfield  {author} {\bibinfo {author} {\bibfnamefont {P.}~\bibnamefont {Romatschke}}\ and\ \bibinfo {author} {\bibfnamefont {M.}~\bibnamefont {Strickland}},\ }\href {\doibase 10.1103/PhysRevD.68.036004} {\bibfield  {journal} {\bibinfo  {journal} {Phys. Rev. D}\ }\textbf {\bibinfo {volume} {68}},\ \bibinfo {pages} {036004} (\bibinfo {year} {2003})},\ \Eprint {http://arxiv.org/abs/hep-ph/0304092} {arXiv:hep-ph/0304092} \BibitemShut {NoStop}%
\bibitem [{\citenamefont {Romatschke}\ and\ \citenamefont {Strickland}(2004)}]{Romatschke:2004jh}%
  \BibitemOpen
  \bibfield  {author} {\bibinfo {author} {\bibfnamefont {P.}~\bibnamefont {Romatschke}}\ and\ \bibinfo {author} {\bibfnamefont {M.}~\bibnamefont {Strickland}},\ }\href {\doibase 10.1103/PhysRevD.70.116006} {\bibfield  {journal} {\bibinfo  {journal} {Phys. Rev. D}\ }\textbf {\bibinfo {volume} {70}},\ \bibinfo {pages} {116006} (\bibinfo {year} {2004})},\ \Eprint {http://arxiv.org/abs/hep-ph/0406188} {arXiv:hep-ph/0406188} \BibitemShut {NoStop}%
\bibitem [{\citenamefont {Kumar}\ \emph {et~al.}(2018)\citenamefont {Kumar}, \citenamefont {Jamal}, \citenamefont {Chandra},\ and\ \citenamefont {Bhatt}}]{Kumar:2017bja}%
  \BibitemOpen
  \bibfield  {author} {\bibinfo {author} {\bibfnamefont {A.}~\bibnamefont {Kumar}}, \bibinfo {author} {\bibfnamefont {M.~Y.}\ \bibnamefont {Jamal}}, \bibinfo {author} {\bibfnamefont {V.}~\bibnamefont {Chandra}}, \ and\ \bibinfo {author} {\bibfnamefont {J.~R.}\ \bibnamefont {Bhatt}},\ }\href {\doibase 10.1103/PhysRevD.97.034007} {\bibfield  {journal} {\bibinfo  {journal} {Phys. Rev. D}\ }\textbf {\bibinfo {volume} {97}},\ \bibinfo {pages} {034007} (\bibinfo {year} {2018})},\ \Eprint {http://arxiv.org/abs/1709.01032} {arXiv:1709.01032 [nucl-th]} \BibitemShut {NoStop}%
\bibitem [{\citenamefont {Jamal}\ \emph {et~al.}(2017)\citenamefont {Jamal}, \citenamefont {Mitra},\ and\ \citenamefont {Chandra}}]{Jamal:2017dqs}%
  \BibitemOpen
  \bibfield  {author} {\bibinfo {author} {\bibfnamefont {M.~Y.}\ \bibnamefont {Jamal}}, \bibinfo {author} {\bibfnamefont {S.}~\bibnamefont {Mitra}}, \ and\ \bibinfo {author} {\bibfnamefont {V.}~\bibnamefont {Chandra}},\ }\href {\doibase 10.1103/PhysRevD.95.094022} {\bibfield  {journal} {\bibinfo  {journal} {Phys. Rev. D}\ }\textbf {\bibinfo {volume} {95}},\ \bibinfo {pages} {094022} (\bibinfo {year} {2017})},\ \Eprint {http://arxiv.org/abs/1701.06162} {arXiv:1701.06162 [nucl-th]} \BibitemShut {NoStop}%
\bibitem [{\citenamefont {Formanek}\ \emph {et~al.}(2021)\citenamefont {Formanek}, \citenamefont {Grayson}, \citenamefont {Rafelski},\ and\ \citenamefont {M\"uller}}]{Formanek:2021blc}%
  \BibitemOpen
  \bibfield  {author} {\bibinfo {author} {\bibfnamefont {M.}~\bibnamefont {Formanek}}, \bibinfo {author} {\bibfnamefont {C.}~\bibnamefont {Grayson}}, \bibinfo {author} {\bibfnamefont {J.}~\bibnamefont {Rafelski}}, \ and\ \bibinfo {author} {\bibfnamefont {B.}~\bibnamefont {M\"uller}},\ }\href {\doibase 10.1016/j.aop.2021.168605} {\bibfield  {journal} {\bibinfo  {journal} {Annals Phys.}\ }\textbf {\bibinfo {volume} {434}},\ \bibinfo {pages} {168605} (\bibinfo {year} {2021})},\ \Eprint {http://arxiv.org/abs/2105.07897} {arXiv:2105.07897 [physics.plasm-ph]} \BibitemShut {NoStop}%
\bibitem [{\citenamefont {Elias}\ \emph {et~al.}(2014)\citenamefont {Elias}, \citenamefont {Peralta-Ramos},\ and\ \citenamefont {Calzetta}}]{Elias:2014hua}%
  \BibitemOpen
  \bibfield  {author} {\bibinfo {author} {\bibfnamefont {M.}~\bibnamefont {Elias}}, \bibinfo {author} {\bibfnamefont {J.}~\bibnamefont {Peralta-Ramos}}, \ and\ \bibinfo {author} {\bibfnamefont {E.}~\bibnamefont {Calzetta}},\ }\href {\doibase 10.1103/PhysRevD.90.014038} {\bibfield  {journal} {\bibinfo  {journal} {Phys. Rev. D}\ }\textbf {\bibinfo {volume} {90}},\ \bibinfo {pages} {014038} (\bibinfo {year} {2014})},\ \Eprint {http://arxiv.org/abs/1404.7790} {arXiv:1404.7790 [hep-ph]} \BibitemShut {NoStop}%
\bibitem [{\citenamefont {Han}\ \emph {et~al.}(2017)\citenamefont {Han}, \citenamefont {Hou}, \citenamefont {Jiang},\ and\ \citenamefont {Li}}]{Han:2017nfz}%
  \BibitemOpen
  \bibfield  {author} {\bibinfo {author} {\bibfnamefont {C.}~\bibnamefont {Han}}, \bibinfo {author} {\bibfnamefont {D.-f.}\ \bibnamefont {Hou}}, \bibinfo {author} {\bibfnamefont {B.-f.}\ \bibnamefont {Jiang}}, \ and\ \bibinfo {author} {\bibfnamefont {J.-r.}\ \bibnamefont {Li}},\ }\href {\doibase 10.1140/epja/i2017-12400-9} {\bibfield  {journal} {\bibinfo  {journal} {Eur. Phys. J. A}\ }\textbf {\bibinfo {volume} {53}},\ \bibinfo {pages} {205} (\bibinfo {year} {2017})}\BibitemShut {NoStop}%
\bibitem [{\citenamefont {Yousuf~Jamal}\ and\ \citenamefont {Chandra}(2019)}]{YousufJamal:2019pen}%
  \BibitemOpen
  \bibfield  {author} {\bibinfo {author} {\bibfnamefont {M.}~\bibnamefont {Yousuf~Jamal}}\ and\ \bibinfo {author} {\bibfnamefont {V.}~\bibnamefont {Chandra}},\ }\href {\doibase 10.1140/epjc/s10052-019-7278-2} {\bibfield  {journal} {\bibinfo  {journal} {Eur. Phys. J. C}\ }\textbf {\bibinfo {volume} {79}},\ \bibinfo {pages} {761} (\bibinfo {year} {2019})},\ \Eprint {http://arxiv.org/abs/1907.12033} {arXiv:1907.12033 [nucl-th]} \BibitemShut {NoStop}%
\bibitem [{\citenamefont {Ghosh}\ \emph {et~al.}(2023)\citenamefont {Ghosh}, \citenamefont {Jamal},\ and\ \citenamefont {Kurian}}]{Ghosh:2023ghi}%
  \BibitemOpen
  \bibfield  {author} {\bibinfo {author} {\bibfnamefont {R.}~\bibnamefont {Ghosh}}, \bibinfo {author} {\bibfnamefont {M.~Y.}\ \bibnamefont {Jamal}}, \ and\ \bibinfo {author} {\bibfnamefont {M.}~\bibnamefont {Kurian}},\ }\href@noop {} {\  (\bibinfo {year} {2023})},\ \Eprint {http://arxiv.org/abs/2306.10247} {arXiv:2306.10247 [hep-ph]} \BibitemShut {NoStop}%
\bibitem [{\citenamefont {Jamal}\ \emph {et~al.}(2023)\citenamefont {Jamal}, \citenamefont {Prakash}, \citenamefont {Nilima},\ and\ \citenamefont {Bandyopadhyay}}]{Jamal:2023ncn}%
  \BibitemOpen
  \bibfield  {author} {\bibinfo {author} {\bibfnamefont {M.~Y.}\ \bibnamefont {Jamal}}, \bibinfo {author} {\bibfnamefont {J.}~\bibnamefont {Prakash}}, \bibinfo {author} {\bibfnamefont {I.}~\bibnamefont {Nilima}}, \ and\ \bibinfo {author} {\bibfnamefont {A.}~\bibnamefont {Bandyopadhyay}},\ }\href@noop {} {\  (\bibinfo {year} {2023})},\ \Eprint {http://arxiv.org/abs/2304.09851} {arXiv:2304.09851 [hep-ph]} \BibitemShut {NoStop}%
\bibitem [{\citenamefont {Braaten}\ and\ \citenamefont {Thoma}(1991)}]{PhysRevD.44.R2625}%
  \BibitemOpen
  \bibfield  {author} {\bibinfo {author} {\bibfnamefont {E.}~\bibnamefont {Braaten}}\ and\ \bibinfo {author} {\bibfnamefont {M.~H.}\ \bibnamefont {Thoma}},\ }\href {\doibase 10.1103/PhysRevD.44.R2625} {\bibfield  {journal} {\bibinfo  {journal} {Phys. Rev. D}\ }\textbf {\bibinfo {volume} {44}},\ \bibinfo {pages} {R2625} (\bibinfo {year} {1991})}\BibitemShut {NoStop}%
\bibitem [{\citenamefont {Mrowczynski}(1991)}]{Mrowczynski:1991da}%
  \BibitemOpen
  \bibfield  {author} {\bibinfo {author} {\bibfnamefont {S.}~\bibnamefont {Mrowczynski}},\ }\href {\doibase 10.1016/0370-2693(91)90188-V} {\bibfield  {journal} {\bibinfo  {journal} {Phys. Lett. B}\ }\textbf {\bibinfo {volume} {269}},\ \bibinfo {pages} {383} (\bibinfo {year} {1991})}\BibitemShut {NoStop}%
\bibitem [{\citenamefont {Thoma}(1991)}]{Thoma:1991jum}%
  \BibitemOpen
  \bibfield  {author} {\bibinfo {author} {\bibfnamefont {M.~H.}\ \bibnamefont {Thoma}},\ }\href {\doibase 10.1016/0370-2693(91)90565-8} {\bibfield  {journal} {\bibinfo  {journal} {Phys. Lett. B}\ }\textbf {\bibinfo {volume} {273}},\ \bibinfo {pages} {128} (\bibinfo {year} {1991})}\BibitemShut {NoStop}%
\bibitem [{\citenamefont {Hosoya}\ and\ \citenamefont {Kajantie}(1985)}]{Hosoya:1983xm}%
  \BibitemOpen
  \bibfield  {author} {\bibinfo {author} {\bibfnamefont {A.}~\bibnamefont {Hosoya}}\ and\ \bibinfo {author} {\bibfnamefont {K.}~\bibnamefont {Kajantie}},\ }\href {\doibase 10.1016/0550-3213(85)90499-7} {\bibfield  {journal} {\bibinfo  {journal} {Nucl. Phys. B}\ }\textbf {\bibinfo {volume} {250}},\ \bibinfo {pages} {666} (\bibinfo {year} {1985})}\BibitemShut {NoStop}%
\bibitem [{\citenamefont {Puglisi}\ \emph {et~al.}(2014)\citenamefont {Puglisi}, \citenamefont {Plumari},\ and\ \citenamefont {Greco}}]{Puglisi:2014sha}%
  \BibitemOpen
  \bibfield  {author} {\bibinfo {author} {\bibfnamefont {A.}~\bibnamefont {Puglisi}}, \bibinfo {author} {\bibfnamefont {S.}~\bibnamefont {Plumari}}, \ and\ \bibinfo {author} {\bibfnamefont {V.}~\bibnamefont {Greco}},\ }\href {\doibase 10.1103/PhysRevD.90.114009} {\bibfield  {journal} {\bibinfo  {journal} {Phys. Rev. D}\ }\textbf {\bibinfo {volume} {90}},\ \bibinfo {pages} {114009} (\bibinfo {year} {2014})},\ \Eprint {http://arxiv.org/abs/1408.7043} {arXiv:1408.7043 [hep-ph]} \BibitemShut {NoStop}%
\bibitem [{\citenamefont {Hongo}\ \emph {et~al.}(2017)\citenamefont {Hongo}, \citenamefont {Hirono},\ and\ \citenamefont {Hirano}}]{Hongo:2013cqa}%
  \BibitemOpen
  \bibfield  {author} {\bibinfo {author} {\bibfnamefont {M.}~\bibnamefont {Hongo}}, \bibinfo {author} {\bibfnamefont {Y.}~\bibnamefont {Hirono}}, \ and\ \bibinfo {author} {\bibfnamefont {T.}~\bibnamefont {Hirano}},\ }\href {\doibase 10.1016/j.physletb.2017.10.028} {\bibfield  {journal} {\bibinfo  {journal} {Phys. Lett. B}\ }\textbf {\bibinfo {volume} {775}},\ \bibinfo {pages} {266} (\bibinfo {year} {2017})},\ \Eprint {http://arxiv.org/abs/1309.2823} {arXiv:1309.2823 [nucl-th]} \BibitemShut {NoStop}%
\bibitem [{\citenamefont {Satow}(2014)}]{Satow:2014lia}%
  \BibitemOpen
  \bibfield  {author} {\bibinfo {author} {\bibfnamefont {D.}~\bibnamefont {Satow}},\ }\href {\doibase 10.1103/PhysRevD.90.034018} {\bibfield  {journal} {\bibinfo  {journal} {Phys. Rev. D}\ }\textbf {\bibinfo {volume} {90}},\ \bibinfo {pages} {034018} (\bibinfo {year} {2014})},\ \Eprint {http://arxiv.org/abs/1406.7032} {arXiv:1406.7032 [hep-ph]} \BibitemShut {NoStop}%
\end{thebibliography}%

\end{document}